\documentclass[letterpaper,twocolumn,10pt]{article}
\usepackage{usenix}

\newcommand\marklessfootnote[1]{
    \addtocounter{footnote}{1} %
    \footnotetext{#1}
}
\usepackage{graphicx}
\usepackage{xcolor}
\usepackage{xspace}
\usepackage{enumitem}
\usepackage{tikz}
\usepackage{subcaption}

\usepackage{hyperref}
\usepackage{url}

\usepackage{breakurl}
\usepackage{balance}
\usepackage{booktabs}
\usepackage{multirow}
\usepackage{adjustbox}
\usepackage{colortbl}

\usepackage{siunitx}
\usepackage{titlesec}
\usepackage{makecell}
\usepackage{verbatimbox}
\usepackage{ragged2e}
\usepackage{tabularx}

\newcommand{\tableappendixspacing}{}%

\newcommand{\codename}{{\sc Acai}\xspace}
\newcommand{\codenameopt}{{\sc Acai\textsubscript{opt}}\xspace}

\renewcommand{\paragraph}[1]{{\bf{\noindent #1}} }
\usepackage{todonotes}

\newcommand{\nss}{protected region\xspace}
\newcommand{\tfa}{TF-A\xspace}

\newcommand{\loc}{LoC\xspace}

\newcommand{\rmidatacreate}{$\tt{rmi\_data\_create}$\xspace}
\newcommand{\rmigranuledelegate}{$\tt{rmi\_granule\_delegate}$\xspace}
\newcommand{\smcdeviceattach}{$\tt{smc\_device\_attach}$\xspace}

\newcommand{\smcdelegateprotmem}{$\tt{smc\_delegate\_prot\_mem}$\xspace}
\newcommand{\rsidelegateprotmem}{$\tt{rsi\_delegate\_prot\_mem}$\xspace}

\newcommand{\attachdev}{$\tt{attach\_dev}$\xspace}
\newcommand{\world}{$\tt{world}$\xspace}
\newcommand{\worldext}{$\tt{world\_ext}$\xspace}

\newcommand{\coregpt}{GPC\textsubscript{c}\xspace}
\newcommand{\devgpt}{GPC\textsubscript{d}\xspace}

\SetEnumitemKey{myenum}{leftmargin = *}

\newlist{invenum}{enumerate}{1}
\setlist[invenum]{label*=INV\arabic*:~,ref=INV\arabic*}
\makeatletter
\newcommand\myitem[1][]{%
  \if\relax\detokenize{#1}\relax
    \item\relax
  \else
    \protected@edef\@currentlabel{\textcolor{black}{INV$_{#1}$}} %
    \item[\textbf{INV$_{#1}$:}]
  \fi}
\makeatother

\newcommand{\bvan}{$B_{v}$\xspace}
\newcommand{\avan}{$A_{v}$\xspace}
\newcommand{\brl}{$B_{r}$\xspace}
\newcommand{\aenc}{$A_{e}$\xspace}
\newcommand{\aprot}{$A_{p}$\xspace}
\newcommand{\aopt}{$A_{o}$\xspace}

\newcommand{\pcie}{PCIe\xspace}
\newcommand{\pciefive}{PCIe-5\xspace}

\newcommand{\middlespacesave}{}%

\newcommand{\spacesave}{}%
\newcommand{\lessspacesave}{}%
\newcommand{\paraspacesave}{}%

\newcommand{\circled}[1]{\textcircled{\raisebox{-0.9pt}{#1}}}

\begin{document}
\date{}

\title{\codename: Protecting Accelerator Execution \\with Arm Confidential Computing Architecture }

\author{\rm Supraja Sridhara \; Andrin Bertschi \; Benedict Schlüter\; Mark Kuhne \; Fabio Aliberti \; Shweta Shinde  \\
{ETH Zurich}}

\maketitle
\begin{abstract}

\marklessfootnote{Extended version of the  Usenix Security 2024 paper.}

Trusted execution environments in several existing and upcoming CPUs demonstrate the success of confidential computing, with the caveat that tenants cannot securely use accelerators such as GPUs and FPGAs.
In this paper, we reconsider the Arm Confidential Computing Architecture (CCA) design, an upcoming TEE feature in Armv9-A, to address this gap.
We observe that CCA offers the right abstraction and mechanisms to allow confidential VMs to use accelerators as a first-class abstraction.
We build \codename, a CCA-based solution, with a principled approach of extending CCA security invariants to device-side access to address several critical security gaps.
Our experimental results on GPU and FPGA demonstrate the feasibility of \codename while maintaining security guarantees.
\end{abstract}

\pagestyle{plain} %

\spacesave
\section{Introduction}

Confidential computing allows cloud users to deploy security-sensitive applications on an untrusted cloud provider's platform. The trusted hardware provisioned in the cloud guarantees that the user's code and data are protected from malicious tenants as well as privileged administrators. 
While some hardware support for enabling trusted execution environments, such as enclaves in Intel SGX, is limited to process-level abstractions, there has been an increasing shift towards VM-level isolation of {\em confidential VMs}~\cite{arm-da, tdx, amd-sev}.
Another noticeable shift, especially in the cloud setting, is towards using accelerators (e.g., GPUs, FPGAs, TPUs) to meet the demands of large workloads~\cite{forbes-gpt, azure-gpt}.
However, trusted execution on CPUs does not extend to accelerators. This leaves cloud users with a dilemma: render the workloads open to attacks by executing them outside the trusted execution boundary or take a performance hit by executing these workloads on CPUs. 

Arm announced specification to enable 
confidential computing architecture, Arm CCA for short, in 2021~\cite{arm-da}. 
CCA isolates confidential VMs, referred to as realm VMs, from each other and the hypervisor with hardware-level access control via a newly introduced mechanism called granule protection checks. 
More importantly, these checks extend beyond the CPU cores to bus transactions as well as address translation for I/O operations, thus allowing CCA-enabled peripherals to securely access VMs. 
Further, CCA's VM abstraction is a good fit for the cloud setting that reasons about 
VMs and accelerators at a coarse granularity. For example, a P3 or F1 instance is provisioned as a bundle of a VM with a dedicated GPU or an FPGA respectively. 
Moreover, the accelerators are connected as PCIe devices.
This is unlike mobile or desktops with integrated accelerators which are used intermittently by the applications (e.g., a browser renders a GUI with a GPU which is then used by a game app).

At the outset, accelerating confidential computing may seem trivial; 
especially in cases where accelerators, such as a GPU, have device-level TEEs.
Accelerators require hardware modifications to enable TEE operations, which requires redesigning the device operations without loss of performance and increase in area size, as showcased with recent Nvidia GPUs~\cite{nvidia-azure}.
Despite these hardware changes, we observe two main challenges remain.
First, the accelerators cannot reason about the security of the host device, such as the authenticity of a realm VM, it connects to. 
Second, on the CPU side, the TEEs have to account for secure device access. 
The realm VMs have to be protected from rogue devices as well as devices controlled by co-tenant VMs. 
Achieving end-to-end secure accelerator execution requires addressing these gaps. 

In this paper, we reconsider Arm CCA VMs with the goal of making access to accelerator devices a first-class primitive. 
Our choice of Arm CCA is inspired partly because it is an upcoming extension, unlike AMD SEV and Intel TDX which are already in production. 
The ability to experiment with CCA on Arm's Fixed Virtual Platform 
allows us to showcase the design choices and reason about its security transparently.
CCA's native support for bus-level access control and memory protection allows us to re-purpose existing hardware protection mechanisms to our setting of PCIe peripherals. 
To this end we present \codename, a CCA-based design that enables realm VMs to securely access accelerators.

Enabling \codename requires careful consideration of security and compatibility implications. 
In the existing CCA specification, external accelerators connected over PCIe can be allowed to access realm memory.
If allowed to do so, such accelerators can break CCA protection that requires systematic enforcement of granule protection checks. 
Next, \codename has to ensure that accesses from different accelerators are isolated to their designated share connected to the appropriate VM while thwarting accesses from malicious peripherals. 
Further, \codename has to isolate all possible I/O and memory access paths, DMA and memory-mapped IO, from software and physical adversaries. Specifically, the hypervisor can manipulate the hardware I/O subsystem and a physical adversary can snoop and tamper the PCIe communication to bypass CPU and device TEE protections.
Lastly, \codename has to maintain compatibility i.e., incur minimal changes to the trusted RMM, monitor and guest Linux kernel as well as the untrusted hypervisor; especially while retrofitting software stacks written with a hierarchical trust model to trusted execution model.

Our main insight is to extend Arm CCA's security invariants for CPU-side protection to reason about devices. 
When a CPU core accesses a memory address, CCA checks if the CPU's execution mode is allowed to access the mode of the memory being accessed, thus protecting the realm from privileged software adversaries (OS and Hypervisor). \codename enables a device to be securely marked as realm-mode, to leverage world-level isolation. 
CCA further isolates realm memory between VMs by using the MMU controlled by trusted software. \codename uses a device-side MMU, called the SMMU, to similarly achieve VM-level isolation but from device accesses. While doing so, \codename ensures that an untrusted hypervisor who can program the SMMU cannot tamper with the isolation guarantees.
CCA employs bus-level memory encryption to protect VMs from physical adversaries, \codename re-uses native hardware-level encryption support from PCIe and the device-side TEE to protect device-bound communication. 
CCA binds the VM's identity and encryption keys to the attestation report. \codename includes the device attestation in the VM creation flow to additionally bind the device identity to the realm VM.
In summary, \codename invariants allow us to systematically reason about the confidentiality and integrity of the VM and device execution. 

We prototype \codename on Arm's public simulator that supports CCA.
Arm provides a trusted firmware called TF-A and a trusted hypervisor called realm management monitor, along with an untrusted hypervisor in the form of a patched KVM. 
\codename modifies these three components to enable secure accelerator execution. 
Arm provides a patched Linux Kernel that can execute both in the realm  VM as well as the untrusted host VM.
For compatibility, \codename adapts the kernel and existing device drivers such that confidential user applications can continue to access the accelerators as they would in a legacy setting.
We showcase \codename with two existing accelerators, a GPU and an FPGA.
Our experiments with existing benchmarks and applications show that \codename supports accelerators out of the box, while maintaining compatibility with existing applications. 
When compared to execution without TEE protection on the CPU and the accelerator, \codename incurs $43.5\%$ and $12.1\%$ overhead on average for GPUs and FPGAs respectively.
We confirm that \codename does not slow down the execution of the rest of the system, it incurs a minimal overhead of $3.8\%$ and $1.9\%$ when using the GPU and FPGA.
Since CCA support is not available in CPUs, our performance estimates are based on a simulator and an Armv8 board port. 

\paragraph{Contributions.}
The main contributions of this paper are:
\begin{itemize}[leftmargin=*]
\item 
We present \codename, the first system to demonstrate secure PCIe device access for Arm CCA-based confidential VMs.
\item 
\codename's novel security invariants identify security gaps in  Arm CCA and achieve secure peripheral execution without any hardware changes to the Arm CCA-compliant hardware or TEE-enabled accelerators. 
\item  
Our evaluation shows that \codename is feasible while achieving its security and compatibility goals. \codename is open source.\footnote{\url{https://sectrs-acai.github.io/acai/}}
\end{itemize}
\section{Confidential Acceleration}
\label{sec:confidential-acceleration}

\begin{figure*}[ht]
  \centering
    \includegraphics[width=0.99\textwidth]{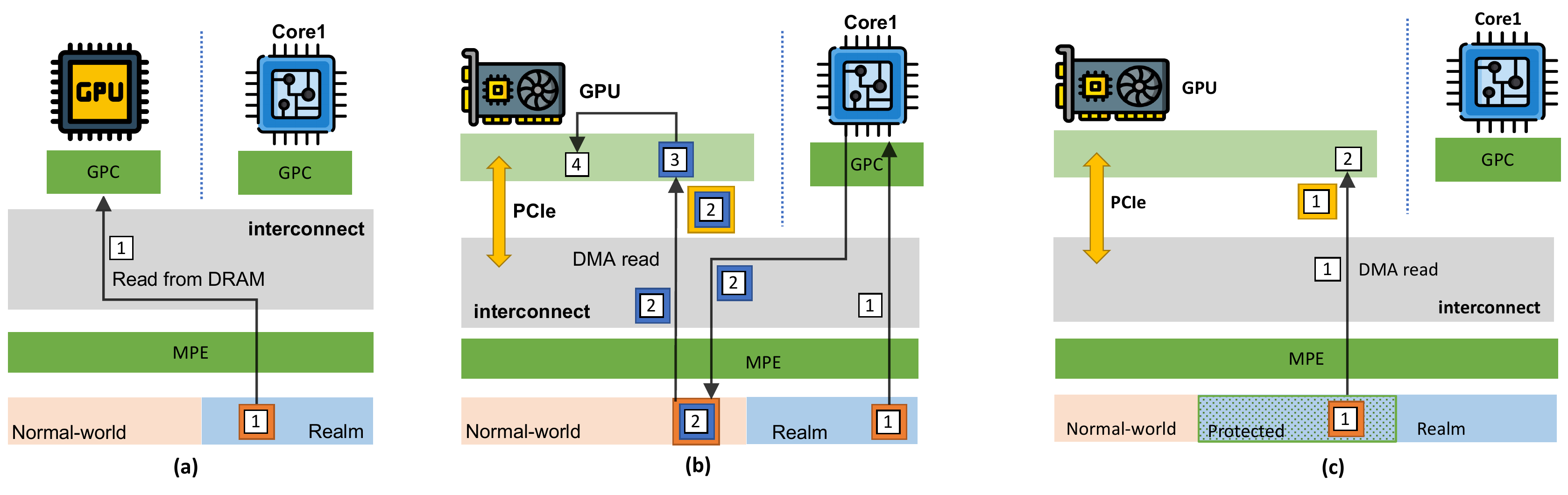}
    \spacesave
    \caption{Access Modes: (a) Integrated (b) Encrypted (c) \codename. The white numbered boxes indicate the original data to be sent to the device, the concentric squares indicate encryption by the CPU bus (orange), software (blue), and PCIe bus (yellow). }
    \label{fig:modes}
\end{figure*}

CCA considers all memory accesses from devices to be 
untrusted by default, thus preventing them from accessing realm memory of confidential VMs. 
Similarly, TEE architectures such as Intel SGX and AMD SEV do not allow devices to access protected enclave and confidential VM memory.

\paragraph{Encrypted Access to PCIe-based Accelerators.}
Most cloud deployments use PCIe to connect powerful CPU cores to dedicated accelerators. This offers the flexibility to plug in the best-suited devices after procurement as well as the scalability to connect $16$-$64$ devices per node. 
The bus-level access control mechanisms, such as Arm TrustZone or RISC-V PMP, do not extend to such devices.
Instead, existing proposals resort to a bounce-buffer design for encrypted communication. The VM encrypts the data buffer in software and sends it to the accelerator, which decrypts it using accelerator-specific logic (e.g., GPU kernel).
Similarly, the accelerator generates results, encrypts them, and transfers them to the publicly accessible part of the CPU memory. Subsequently, the VM has to copy the encrypted data into its own protected memory and then decrypt it.
As shown in Table~\ref{tab:modes} and Figure~\ref{fig:modes}(b), this approach leads to two extra copies and software-based encryption and integrity protection on both the processors and peripherals thus increasing the memory and compute overheads.
This requires invasive API changes in the applications, drivers to perform encryption, and the device-side logic for encryption; thus breaking compatibility.
Case in point, connecting Nvidia GPUs to AMD SEV VMs required cooperation from Nvidia, AMD, and Azure while still achieving 4Gbps bandwidth due to software encryption and memory copy bottlenecks~\cite{nvidia-azure}.

The \pciefive specification introduces IDE (Integrity and Data encryption) which provides confidentiality and integrity guarantees for PCIe packets~\cite{pcie5-IDE-spec}. 
IDE's hardware encryption feature can be leveraged to build a  performant design without the need for software-based encryption. However, IDE is not designed for a threat model with an untrusted hypervisor. 

\begin{table}[]
    \centering
    \caption{Number of data copies and encryption-decryption operations in hardware (Enc\textsubscript{HW}) and software (Enc\textsubscript{SW}) required per transaction between the CPU and the device.}
    \label{tab:modes}
        \resizebox{0.8\columnwidth}{!}{%
        \begin{tabular}{ccc|ccc}
        \hline
        \multicolumn{3}{c|}{Bounce Buffer} &
          \multicolumn{3}{c}{\codename} \\
        Copies &
          Enc\textsubscript{SW} &
          Enc\textsubscript{HW} &
          Copies &
          Enc\textsubscript{SW} &
          Enc\textsubscript{HW} \\ \hline
        3 &
          2 &
          2 &
          1 &
          0 &
          2 \\ \hline
        \end{tabular}%
        }
    \end{table}
\spacesave

\paragraph{Protected Access to Integrated Accelerators.}
In Arm architecture, accelerator IPs can be embedded as part of the SoC. 
Such integrated devices do not have their own main memory; instead, they access the main memory along with the CPU cores.
Most popular examples are Mali and Immortalis GPUs, and Mali NPUs.
Since these components operate over a shared bus that connects to main memory, an access control mechanism that operates at bus-level for CPU isolation can be applied to integrated accelerators as well.
This approach has been previously applied to RISC-V~\cite{cure, pie} and Arm TrustZone~\cite{strongbox}.
The success of this approach relies on two requirements---all device accesses strictly pass through the bus and the device enforces its own access control---that render the device access to be equivalent to a CPU core access for all purposes. 
 
In the context of CCA, a CPU core is allowed to access realm memory if it has the right privileges. 
Specifically, each core can program the access mode (root/realm/secure/normal) for physical address ranges.
The architecture introduces a special data structure, called granule protection table (GPT), to track these access modes. The GPT resides in the root mode memory and can only be accessed by the monitor.
The architecture then converts the GPT restrictions into access checks via granule protection checks (GPCs). 
The GPCs can then be applied to components, such as CPU cores, caches, TLBs, and SMMU, that access the memory.
Whenever there is a change in the GPT, the GPCs are synchronized and may trigger the flushing of stale states. 
Lastly, a physical attacker cannot tamper with the data on the main memory because of bus-level encryption and integrity protection mechanisms.
To allow devices to access realm memory, CCA necessitates that the device implements a device-side GPC, such that any accesses coming from the device is filtered as per the GPT before it is allowed to propagate on the bus.
Figure~\ref{fig:modes}(a) summarizes this mode and Table~\ref{tab:modes} shows that the overheads for this mode are minimal---there are no data copies and the encryption is performed transparently and is at line-rate.

\section{Background: Isolation in Arm CCA}
\begin{figure}[]
    \centering
     \includegraphics[scale=0.37]{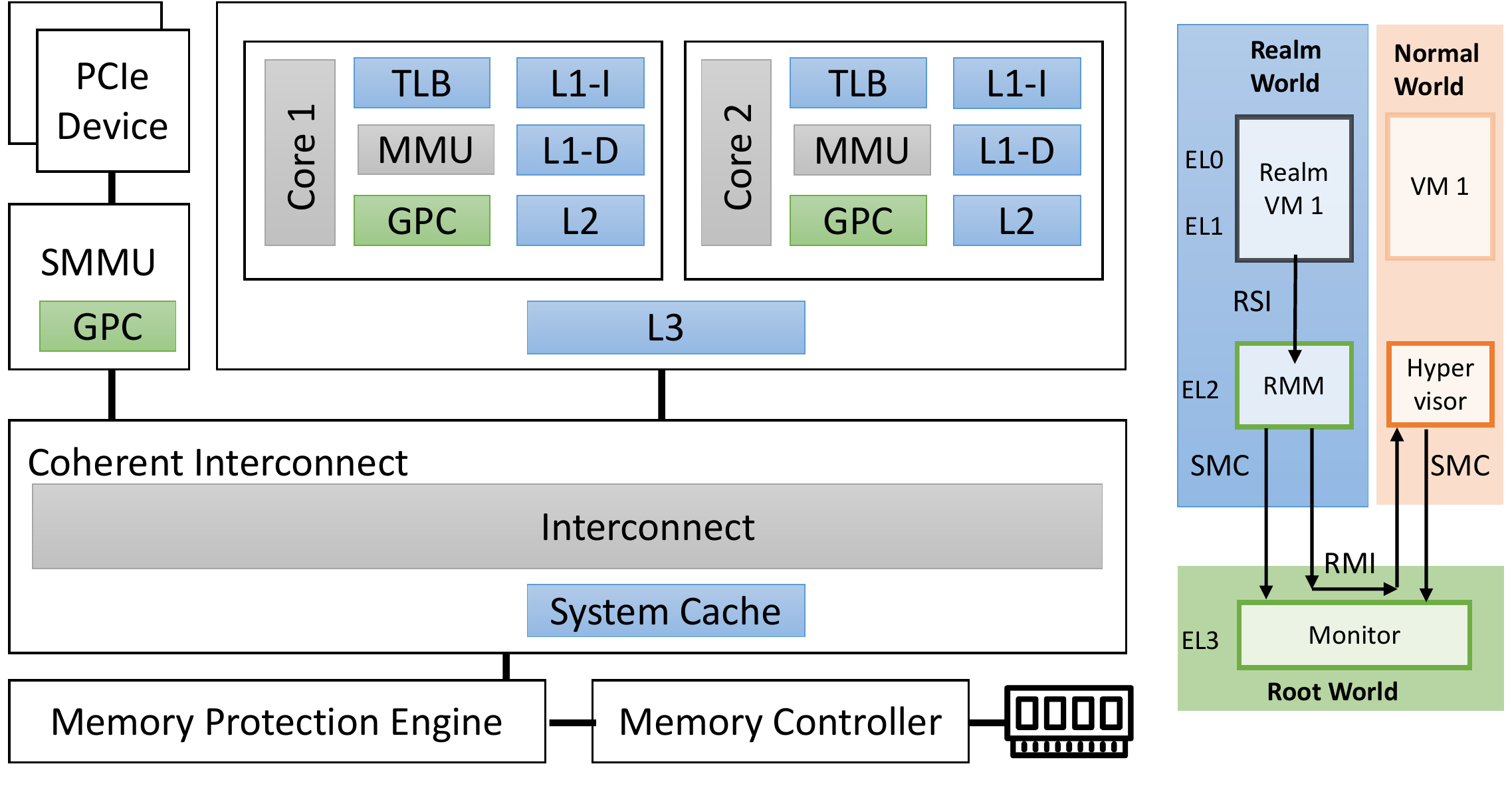}
     \caption{GPC in Arm CCA. Blue components are subject to GPC enforcement. The right figure shows interfaces between the realm VM, RMM, hypervisor, and monitor. Realm service interface (RSI) and realm management interface (RMI) are used by the Realm VM and hypervisor respectively.
     } 
     \vspace{6pt}
     \label{fig:cca-detailed}
 \end{figure}

\spacesave
Arm Confidential Computing Architecture (CCA) enables the creation of VM-based isolated execution environments. 
CCA is enabled by a hardware extension to the Arm ISA, Realm Management Extensions (RME), that creates $2$ new worlds (Realm and Root) along with the $2$ existing worlds (Normal and Secure).
Figure~\ref{fig:cca-detailed} shows how the schematic of exception levels and worlds works when CCA-based VMs are executing. 
On a CCA-enabled platform, a CPU core is allowed to access realm memory if it is in the right privilege and mode. 
Specifically, each core can program the access mode of physical address ranges to be root, realm, secure, or normal world.
The architecture introduces a special data structure, called granule protection table (GPT) that tracks which physical address belongs to which of the four worlds.
To perform the world isolation, CCA augments all processing elements (e.g., Arm cores, SMMU, caches, TLBs) with Granule Protection Checks (GPCs), where a Granule is the smallest block of memory that can be described.
The GPTs are maintained in the main memory and belong to the root world such that only the monitor can access and update them. 
Whenever there is a change in the GPT, the GPCs are synchronized and may trigger the flushing of stale states. 
The GPCs, generated from the GPTs, effectively prevent accesses from the normal or secure world to the realm memory by performing access control on all memory accesses. Specifically, all bus transactions initiated from the core are guarded.
The trusted firmware, executing in the monitor, sets the world bit for the core when it performs a context switch.
Next, a trusted Realm Management Monitor (RMM) executes in the realm world at EL2. 
The RMM ensures that mutually distrusting realm VMs are isolated by managing stage-2 translations from guest physical address
(IPA in Arm terminology) to host physical address (PA in
Arm terminology).
Lastly, a physical attacker cannot tamper with the data when it resides in the main memory because of the bus-level encryption and integrity protection mechanisms.

In summary, CCA enforces isolation for the realm world computation using a combination of stage-2 translations and GPCs. 
However, these isolation techniques do not apply to devices. 
Specifically, when a CPU core accesses memory, it is subject to access control checks by the MMU. 
But, when a PCIe device accesses the host memory, it bypasses the MMU. 
To remediate this, host systems place an SMMU between the device and the memory that can perform access control, such that all PCIe accesses (e.g., GPU requests to DMA data from the host to the device) are subject to checks. 
CCA augments the SMMU's legacy access control with GPC, such that all device accesses to memory are now isolated at a world granularity. 
For example, with TrustZone support, secure world devices are allowed to access secure and normal world memory, whereas normal world devices can only access normal world memory. 
CCA's RME Device Attach extends this notion to the realm world allowing devices to access realm memory. 

\middlespacesave
\section{Security Challenges}
CCA's device-accessible realm memory along with hardware encryption on both CPU and accelerators, points to a performant design that removes the need for multiple data copies and software-based encryption-decryption (see Section~\ref{sec:confidential-acceleration}).
However, doing so requires carefully addressing security considerations to ensure that the end-to-end execution on device and host composes safely. 
We outline the security gaps and potential attacks when attaching \pcie devices to realm VMs. 

\label{sec:overview}
\begin{figure}[]
  \centering
    \includegraphics[scale=0.62]{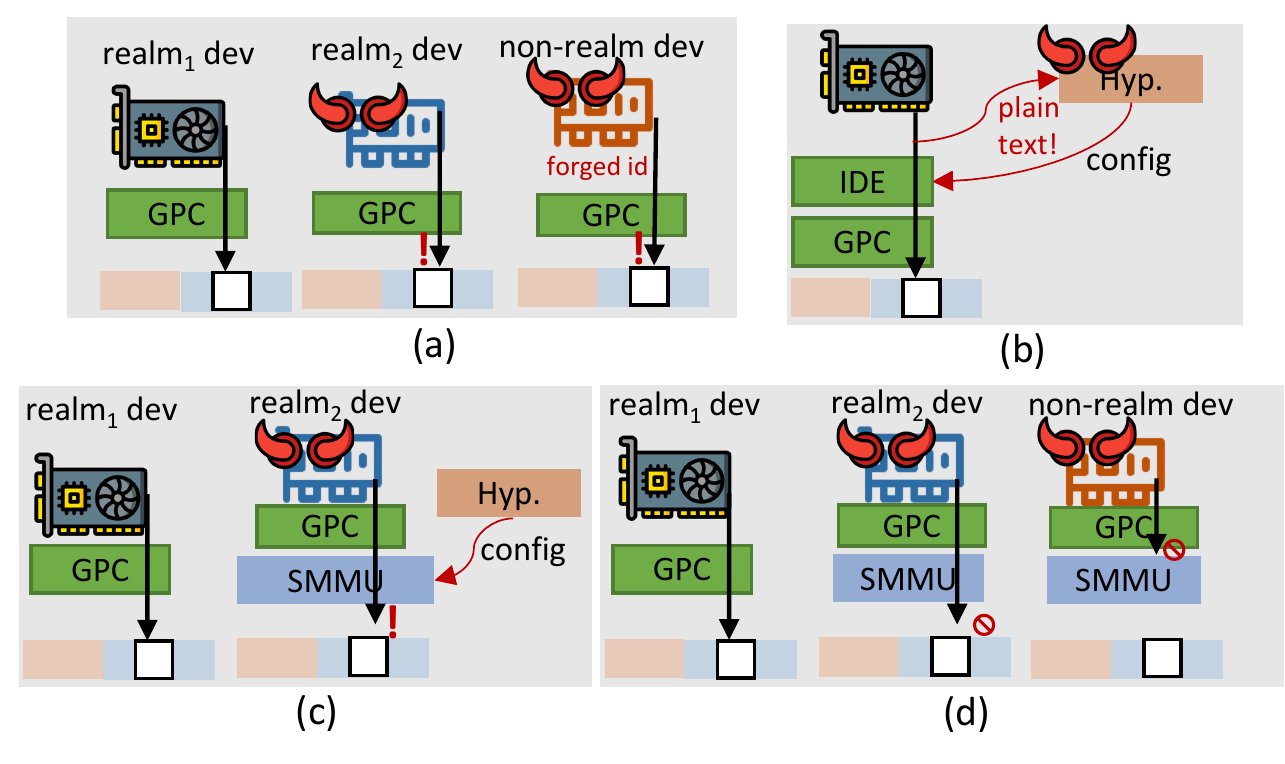}
    \middlespacesave
    \caption{(a)-(c) Potential attacks when devices access realm memory. 
    (d) \codename protection
    }
    \vspace{4pt}
    \label{fig:approach}
\end{figure}
\paragraph{Fine-grain Memory Isolation.}
Devices have to be securely allocated to a realm VM in a way that an attacker cannot launch man-in-the-middle attacks, tunnel the access through untrusted devices, or attach the same device to multiple mutually distrusting realm VMs. 
After device allocation, all its accesses (e.g., DMA) have to be isolated not only to the realm memory but specifically to the realm VM that owns the device. 
Otherwise, malicious devices attached to other realm VMs can compromise computation (Figure~\ref{fig:approach}a). 
Protecting the shared memory between the device and the VM entails stopping other VMs (in the realm or otherwise), devices, and hypervisor from accessing it. 
CCA hardware ensures that the GPCs on the MMU and SMMU are in sync to reflect the latest state of access control rules at world granularity (e.g., that a specific page belongs to the realm or the normal world). For secure device access from a VM, this synchronization has to be extended to be at a VM granularity. 
Note that the isolation guarantees that GPCs provide depend on correctly deciding the world that the device belongs to. 
Therefore, device identities that tie them to worlds must be unforgeable to stop malicious devices from compromising security (Figure~\ref{fig:approach}a). 

\paragraph{Preserving Compatibility with  Untrusted Code.}
Hypervisors, VMs, drivers, and user applications should not require invasive rewriting. 
Compatibility without compromising security requires ensuring that the untrusted hypervisor cannot maliciously reconfigure devices, the MMU, and the SMMU. 
Much like on Intel and AMD, the untrusted hypervisor on Arm controls and programs the SMMU. 
It can mount attacks on realm VMs by changing the SMMU's translation tables to allow an attacker-controlled device to access realm memory (Figure~\ref{fig:approach}c). 
Further, the hypervisor can emulate devices or allocate compromised devices to break the realm VM's data confidentiality or execution integrity. 
Hypervisor optimization techniques that allow multiple VMs or devices to access the same physical page (e.g., memory sharing or copy-on-write) can pose a threat to the isolation of protected memory. 
Several device and SMMU features bypass all access control protections because they were designed for performance. These need to be reexamined under the CCA threat model as well. 

\paragraph{Secure I/O Paths.}
Physical attackers can unplug devices, change their ports, probe the memory, and tamper with bus and PCIe transactions. 
Preventing such attacks requires bus- and PCIe-level encryption with key management that accounts for abrupt disconnections while ensuring that malicious devices, VMs, and the hypervisor cannot abuse it. 
For example, the untrusted hypervisor has access to the IDE keys in \pciefive and can compromise data protected by its encryption (Figure~\ref{fig:approach}b).

\section{Our Approach}
\codename adopts a principled approach to securely attach devices to realm VMs while preserving the existing CCA guarantees. 

\subsection{\codename Isolation Principle}
\codename sets up a shared memory region between the realm VM and the corresponding device, that we refer to as {\em protected region}, in the realm world memory. 
\codename aims to enforce isolation such that an attacker cannot access the protected region.
Our approach is inspired by CCA's notion of isolation for CPU accesses. 
We use it to explain our design intuition and how it leads to \codename's concrete invariants.

On the CPU, Arm CCA's secure boot and attestation mechanism ensures that the cores are in the right state to run confidential realm VMs. 
Similar to CCA, \codename ensures that the device and the realm start in a clean state and that their identities are uniquely mapped to their dedicated protected memory region (\ref{inv:1}). 
Once the identities are established, \codename has to ensure that the device is uniquely assigned to a realm VM which can configure and manage it. 
Therefore, \codename guarantees that a device and all the memory on the host used to configure the device is uniquely mapped to a realm VM (\ref{inv:2}). 
Further, for physical attacker protection, \codename leverages CCA's memory protection engine such that only the device and the owner VM have the key to encrypt and decrypt the data in the protected region.  

With identity and ownership established, \codename has to ensure the translations from guest PA to host PA used for the protected memory regions are secure and compatible with CCA. 
First, CCA dictates that for any host physical address ${P_h}$ that is allocated to the realm VM, one and only one guest physical address ${P_g}$ can map to it. The RMM, who is in charge of managing the realm world memory, enforces this invariant for the stage-2 translations between ${P_g}$ and ${P_h}$. 
Since \codename extends a realm VM to allow access to the protected memory region, our main intuition is to extend the existing CCA invariant to this protected memory (\ref{inv:4}). 
Next, since the device also has access to the protected memory, \codename ensures that a similar invariant (\ref{inv:5}) is applied to any device-side translations i.e., the SMMU. 
\ref{inv:4} and \ref{inv:5} enforce correct translations for the host and device respectively. 
However, one problem still remains. 
If the host and device-side translations are not synchronized, an attacker can mount split-view type of attacks. 
Therefore, \codename ensures that the realm VM and device always have the same view of memory by ensuring that the RMM and SMMU stage-2 translation tables are always synchronized (\ref{inv:3}).

\paragraph{\codename Invariants.}
\codename aims to achieve the same security guarantees for device-side accesses that Arm cores, monitor, and the RMM achieve for CPU-side accesses. 
More precisely, \codename's design aims to enforce the following invariants:
\begin{invenum}[labelindent=0pt,labelwidth=0.75em,leftmargin=!, wide=0pt]
    \setlength\itemsep{-0.3em}
    
    \myitem[id]\label{inv:1} {\em Identity.} An accelerator attached to a realm VM is assigned a unique unforgeable identity after attestation. 
    \myitem[own]\label{inv:2} {\em Exclusive Ownership.} An accelerator is only accessible to one owner realm VM who is allowed to configure the device and setup protection keys and communicate with it.
    \myitem[bind]\label{inv:3} {\em Owner binding.} Owner realm VM and the accelerators have the same guest PA to host PA mappings.
    \myitem[host]\label{inv:4} {\em 1:1 host and guest page mappings.} 
    Only one VM-level guest PA resolves to the same host PA. 
    \myitem[dev]\label{inv:5} {\em No overlapping physical memory in the protected region.} For each host PA in the protected region, only one accelerator has a valid mapping from guest PA to host PA.
\end{invenum}

\paragraph{Threat Model.}
\codename assumes a malicious cloud provider, co-tenants, and privileged software such as the hypervisor and secure world software
who want to compromise the confidentiality and integrity of a victim's realm VM and its device-bound operations.
A physical attacker can plug in malicious devices, compromise the connectors between nodes, and launch probe attacks on DRAM chips that are not integrated into the devices and SoCs. 
\codename trusts the manufacturer of Arm cores, devices, and PCIe port implementations. 
We include the firmware and the RMM in the TCB and rely on CCA attestation mechanisms to ensure that it is untampered. 
Further, each realm VM's TCB includes the guest OS, device drivers, runtimes necessary to access the devices, and the device-side programmable logic such as kernels and IPs.
The TCB may be prone to Iago attacks when the VMs invoke untrusted hypervisor API. \codename does not add any new API to the hypervisor and relies on existing CCA protection for sanitization and preventing Iago attacks~\cite{checkoway2013iago}. 

\paragraph{Scope.}
\codename does not protect against denial-of-service from the malicious hypervisor or cloud provider but ensures that realm VMs cannot cause denial of service to each other and to the cloud provider. 
Defending against speculation attacks and side-channel leakage due to micro-architectural implementation is orthogonal to our work and \codename relies on existing techniques and CCA's native protection for such attacks. 

\subsection{Creating Protected Memory Regions}
\paragraph{Background.} In Arm CCA, the normal world Hypervisor creates the realm VM (source granules) and
then transitions it to the realm world (destination granules).
This invokes the trusted monitor who updates the GPT to change these pages from
normal to realm world. 
The GPT changes trigger flushing of TLBs and caches of all GPCs.
Then the RMM checks if the destination granule is valid and if so updates its
stage-2 translation tables; thus allowing the realm VM's IPA to translate to the granule's host physical address (PA).

\paragraph{Shared Protected Region.}
\codename requires a shared protected memory region between the realm VM and the accelerator. 
RME-DA enables devices to directly access realm world memory. 
Therefore, the \codename protected memory region can reside in the realm world as shown in Figure~\ref{fig:approach}(d).
\codename has to ensure that only the designated device and realm VM can access it in the presence of malicious devices and VMs (Figure~\ref{fig:approach}d).
\codename uses the RMM page tables to ensure that only the realm VM that owns the protected region has the mappings in RMM's stage-2 translation tables (\ref{inv:4}).
Similar to RMM, \codename programs the SMMU page tables to ensure that only the accelerator that was mapped to the protected region has the entries in the SMMU's stage-2 translation tables (\ref{inv:5}). 
Finally, on any changes to the RMM's page tables corresponding to the protected memory, \codename explicitly synchronizes the SMMU to reflect these changes (\ref{inv:3}). 

\paragraph{Protecting the SMMU.}
In CCA, the hypervisor has direct access to the SMMU by default and
can tamper with \codename enforced mappings (\ref{inv:3}, \ref{inv:5}) as shown in Figure~\ref{fig:approach}c. 
To address this gap, \codename uses the GPT to mark the SMMU data structures (stream tables, stage-2 translation tables per device, request-response buffers) to be only accessible in root mode.
If the hypervisor attempts to program the SMMU, it will fault and trigger a handler in the monitor. 
Moving all the SMMU management code from the hypervisor to the monitor 
would result in a bloated monitor and cause loss of optimizations employed by different hypervisors. Instead, \codename introduces an SMMU interface for the hypervisor to request SMMU updates. 
\codename only allows selective hypervisor requests that satisfy its invariants on all SMMU operations. 

\spacesave
\subsection{Identity and Exclusive Ownership}
\paragraph{Background.} 
PCIe transaction packets contain a RequesterID that identifies the device that originated the packet. 
But devices can set this field to arbitrary values. 
Integrity and Data Encryption (IDE) in \pciefive ensures that each PCIe link per device has a unique key which is used for data protection. 

\paragraph{IDE-based Identity \& Enforcing Exclusive Ownership.}
To create a realm VM, the hypervisor sets up a new VM, prepares the accelerator, and then invokes the monitor. 
Here, \codename amends the CCA's VM creation in the monitor to
simultaneously initiate a secure accelerator attachment to enforce invariants \ref{inv:1} and \ref{inv:2}. 
The monitor probes the bus to locate the accelerator and check 
if the accelerator is not already mapped to any realm VM. If not, the monitor sends a reset signal and requests an attestation report. 
\codename uses the RMM and monitor to ensure that only the owner realm VM has access to the accelerator's configurations ensuring \ref{inv:2}. 
\codename uses IDE to ensure that the accelerator has a unique ID that does not overlap with existing devices (\ref{inv:1}).
Specifically, \codename relies on the IDE keys programmed in the root port of the host to ensure that accelerator IDs are unique. Packets with forged accelerator IDs (e.g., using a malicious device, plugging in a compromised device after attestation) do not decrypt successfully and are discarded at the PCIe root port. 
\codename can use native device attestation support or the high-level building blocks of \pciefive
compliance that can attest the accelerator and sign it with root port
keys~\cite{pcie5-IDE-spec, pcie_attest, arm-da}.
\codename requires that the attestation report from the device contains information about its firmware, configuration, and other state after secure boot. 
The monitor forwards the accelerator report to the RMM, who combines it with the realm VM's attestation for the remote verifier checks.

\section{Accelerator Lifecycle}
\label{ssec:device-initialisation}
\paragraph{PCIe Devices \& Memory.} 
All PCIe devices, which are always connected to the PCIe root port, are behind the SMMU.
To discover PCIe devices, the hypervisor invokes the firmware to probe the bus
and read the configuration and Base Address Registers (BARs) size from the
device. 
Each device can have up to $6$ BARs in its PCIe configuration space. The device
uses the address in the BAR to access regions that are memory-mapped by the
kernel or hypervisor. 
Therefore, a device's memory comprises its PCIe configuration space, the memory-mapped BAR regions for MMIO communication, and dynamically allocated DMA regions. \codename ensures that all these regions are protected according to its invariants.

\paragraph{Device Attach.} \codename augments the realm creation process to attach an accelerator while enforcing~\ref{inv:1} and~\ref{inv:2}. 
Note that attaching an accelerator to a realm VM is similar to adding data to the realm VM in the form of memory-mapped BAR regions and the PCIe configuration space.
So, \codename has to ensure that~\ref{inv:4} is enforced for these memory regions. 

By default, the hypervisor resets the accelerator, disables all debug configurations, and sets up its PCIe configuration space. 
In CCA's realm creation process, the hypervisor moves all memory and data to the realm world before starting the realm VM.
Similarly, before starting such an accelerator attach process, \codename requires that the hypervisor has moved all BAR regions of the accelerator and the PCIe configuration space to the realm world and attached them to the realm VM.\footnote{The data in the PCIe configuration space and BAR regions is deterministic on device reset}
This ensures that the accelerator is locked and not accessible to the hypervisor when it starts the device attach process.
The RMM ensures that the hypervisor has performed these steps faithfully (\ref{inv:4}) before starting the device attach process. 
Locking accelerator access from the hypervisor is important to correctly enforce \ref{inv:1} and \ref{inv:2}. 
Further, attaching the configuration space to the realm VM ensures~\ref{inv:2}.

To attach an accelerator to a realm VM in \codename, 
the hypervisor adds two accelerator-specific details to the realm VM: the contents of the PCIe configuration space that it setup and the size of each BAR region.
The RMM ensures \ref{inv:4} for the PCIe configuration space. It checks that the IPA to PA mapping for the PCIe configuration space is valid and that the PA is in realm world. 
PCIe BARs represent contiguous regions of memory of sizes as indicated by the hypervisor. 
To ensure that the memory-mapped regions of the accelerators are correctly mapped to the realm VM (\ref{inv:4}), for every BAR in the PCIe configuration space, the RMM checks that the whole memory region is already mapped to the realm VM. 

\paragraph{Initialization \& Device Attestation.} 
In \codename, the monitor initializes the device and sets it up to ensure~\ref{inv:1}. 
The RMM forwards the PCIe accelerator's bus address that it receives from the hypervisor and the PCIe configuration region's physical address to the monitor.
The monitor uses the accelerator's bus address to re-scan the PCIe bus, 
proceeds to read the accelerator's configuration, and writes that to the physical address of the PCIe configuration space sent by the RMM. 

\codename monitor implements the attestation protocol according to the Security Protocols and Data Models (SPDM), a standard to enable authentication and attestation which can be used for devices~\cite{SPDMattestation, arm-da}. 
As part of the SPDM protocol, the monitor also invokes the IDE key management protocol according to the PCIe specifications~\cite{pcie5-IDE-spec}.
The monitor blocks the return of this call until it receives the attestation report from the accelerator. 
When the protocol completes and the monitor receives the attestation report, it writes the attestation report to the RMM memory 
and returns from the call back to the RMM. 
Note that \codename performs IDE key programming before reading the configuration and attestation reports from the accelerator. This prevents TOCTOU attacks---if the accelerator is replaced or reset after the key programming, all communications from the host will be encrypted with a key that is not accessible to the device.
Therefore, this process securely resets the accelerator, creates an unforgeable identity using IDE, and collects the attestation report. 

\paragraph{Realm Attestation.}
\codename extends the attestation process of CCA to enable a remote verifier to check the attestation reports from the device. 
In CCA, the RMM extends the attestation measurement over the data that was copied into the realm VM. 
\codename reuses this mechanism in the RMM for the memory-mapped regions and PCIe configuration space. 
Specifically, the BAR regions mapped to the device are added to the realm VM like any other realm VM memory. 
Further, the bus address of the accelerator, and the PCIe configuration space including the size of the BAR regions is part of the data added to the realm VM by the hypervisor.
So the RMM's attestation report will indicate the accelerator attached to the realm VM, the PCIe configuration space, and the regions of memory that are used for the BARs and their sizes. 
Further, the RMM appends a new section to the attestation report with the accelerator attestation report that is returned by the monitor.
A remote verifier can now use the attestation report to ensure that the accelerator is correctly initialized (e.g.,  correct firmware, debug disabled, non-PCIe ports disabled), attested, and all device-mapped memory is in the protected region.
 
\paragraph{Establishing Secure Accelerator Identities.}
In PCIe-5 with Integrity and Data Encryption (IDE), the data on the PCIe link is protected from a physical adversary. 
\codename relies on PCIe features to establish unique accelerator identities (\ref{inv:1}).
The IDE keys are stored in the physical root ports of PCIe endpoints, preventing a software adversary such as the hypervisor from tampering or replacing them.
The manufacturer is expected to ensure this, either by keeping the keys in hardware-protected memory in the root port or in the protected SRAM of the host system~\cite{synopses-ide}. 
\codename requires these keys to be stored in memory that is only accessible to the monitor (e.g., root world SRAM).
The PCIe root port is programmed by the monitor with a unique key during the device attach process. 
The monitor ensures that the RequesterID (RID) used for the key mapping is unique. 
If a malicious device originates PCIe transaction packets with forged RIDs, the packets will fail to decrypt in the root port and hence will be discarded. 

\paragraph{Device Detach.}
\codename only allows the hypervisor to detach devices from a realm VM in the realm VM destruction phase. 
When a realm is destroyed, the RMM invokes a call to the monitor to remove any devices attached to the realm with the realm's VMID. 
The monitor removes any device-to-VM mappings it has, and deletes the corresponding device state from the SMMU (see Section~\ref{ssec:secure-smmu-setup}).
All device memory is removed from the realm VM according to normal CCA mechanisms for realm destruction.
Specifically, the hypervisor invokes the RMM to transition the granules back to the normal world. 
In Arm's RMM specification, in this phase, the monitor transitions all realm VM granules back to the normal world after the hardware scrubs the data in the pages. 

\paragraph{Design Decisions.}
We opt for a conservative but secure design where devices have to be attached at realm VM creation and detached at realm VM destruction. This is in line with the cloud model, where the tenant has to pay for a VM and a device attached to it and is not allowed to dynamically allocate accelerators. 
For example, on AWS a VM with a GPU (P3) or an FPGA (F1) instance is provisioned together, as in \codename. 
Some accelerators (e.g., GPUs) can operate in multi-tenant mode where a physical accelerator is shared among multiple VMs.
Enabling such virtual shares in our threat model requires native device virtualization (e.g., MIGs on H100~\cite{mig}) to provide isolation and per-unit encryption keys. If available, \codename is compatible to enable multi-tenancy because the IDE specification and CCA DA support device virtualization.
However, we leave it to future work to closely examine the security challenges of device virtualization. 
For example, virtual shares have common state (e.g., address translation tables, cached translations) that an attacker can exploit.

\section{Memory Isolation}
\label{sec:memory-isolation}
Arm SMMU is an IOMMU that performs access control and address translation for devices using StreamID to identify the devices. 
The hypervisor sets up and manages the SMMU's configuration and stage-1 and stage-2 translations for the normal world.
Specifically, the hypervisor configures the SMMU with the Stream Table's base address. The Stream Table contains entries called Stream Table Entries (STEs) that describe the configuration and translation tables for each device. 
For each incoming transaction, 
the SMMU uses StreamIDs to index into the Stream Table to look up the STE. 
Further, to perform stage-2 translation, the SMMU reads the translation table's base address from the STE. 
Before Arm CCA, the SMMU could differentiate incoming transactions as normal or secure based on a 1-bit long \world bit.
To support realm transactions, CCA extends the \world bit to be 2-bits long (\worldext). 
The SMMU expects the \worldext bits to be set correctly, either by the device or the root port, for all transactions and does not perform any additional checks.

Applications use MMIO and DMA to communicate with the devices. 
They use memory-mapped IO (MMIO) regions to transfer small amounts of data to devices. Physical addresses that can be memory-mapped to devices are not backed by actual host memory (e.g., DRAM). 
Instead, the application directly accesses the device memory using address translation. 
The BARs indicate all the MMIO regions used by the device attached to a realm VM.
In Section~\ref{ssec:secure-smmu-setup} and ~\ref{ssec:mmio-dma}, we explain how \codename protects the SMMU and MMIO/DMA operations.

\spacesave
\subsection{SMMU Changes and Secure Setup}
\label{ssec:secure-smmu-setup}
\begin{figure}
    \centering
     \includegraphics[scale=0.6]{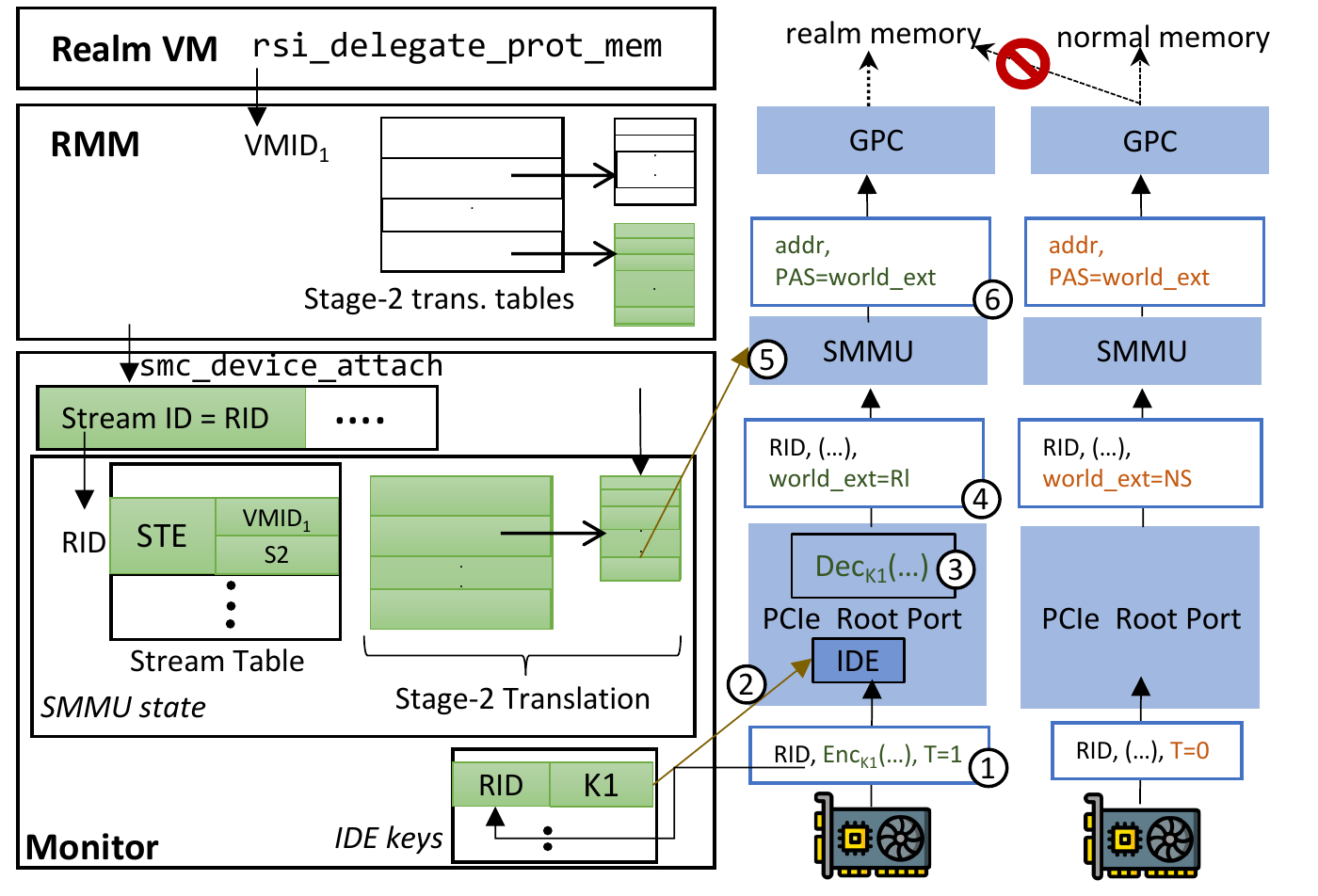} %
         \spacesave
         \caption{\codename's interfaces and software.
         Green: new components; Brown arrows: accesses made with root world PAS.
         \circled{1} accelerator originates transaction with RID and T=1 for realm transactions. The data is encrypted with key K1 setup for IDE. \circled{2} Host Root Port (RP) reads K1 based on RID \circled{3} to decrypt. \circled{4} RP sends transaction to SMMU with \worldext bits set to realm world. \circled{5} based on \worldext, SMMU indexes Realm Stream Table (as root world) with RID to perform stage-2 translation and gets a physical address. \circled{6} SMMU tags the memory address with \worldext for GPC checks.
         } 
         \vspace{4pt}
    \label{fig:smmu-state}
 \end{figure}

\paraspacesave
\paragraph{World-level Isolation.} Arm CCA enables realm world transactions and a realm programming interface in the SMMU~\cite{arm-da}. But it does not explain how one would identify a device to be associated with a realm VM. 
Further, CCA states that the protocol to set the \worldext bits is implementation defined~\cite{arm-smmu-spec}. 
In \codename, identifying and binding transactions to realm VMs is critical for security. 
Therefore,  
\codename defines how to set the \worldext securely to identify realm transactions and bind transactions to realm VMs, as summarized in Figure~\ref{fig:smmu-state}. 

For each core, CCA has a core-specific register to track its execution world. The monitor sets this world for each core on context switch. The GPC reads this register to perform memory access checks. 
Similarly, the GPC on the SMMU needs to identify the world of each memory access from the device. 
\codename leverages PCIe-5 IDE to infer the world of every device transaction to and set the \worldext bits. 
Specifically, it stipulates that all realm world transactions must be subject to IDE.
\codename expects the devices to explicitly tag all realm world transactions by setting a T bit. 
Note that, a transaction with T=0 cannot access realm world memory and will be rejected by the GPC as shown in Figure~\ref{fig:smmu-state}.
Any transaction that accesses memory in the realm world (T=$1$ in Figure~\ref{fig:smmu-state}) has to be successfully decrypted in the PCIe root port. 
Therefore, this mechanism guarantees that the RID and the T bit are correct and can be safely used to infer the world of the transaction ($\tt{world\_ext}$=Rl in Figure~\ref{fig:smmu-state}). 
This approach ensures that the SMMU can correctly perform translation for all devices. 

\paragraph{Intra-realm Isolation.} With the above approach, GPC on the SMMU can isolate realm memory from non-realm devices. 
However, this is not sufficient to isolate mutually distrusting realm devices (Figure~\ref{fig:approach}a). 
The RMM isolates mutually distrusting realm VMs using stage-2 translations. It stores these tables and other realm configurations in protected realm memory inaccessible to the hypervisor. 
Correspondingly, \codename uses the SMMU's stage-2 translation to isolate mutually distrusting devices from accessing each other's protected realm memory.
Updates to the SMMU's stage-2 translation tables must ensure \ref{inv:3} and \ref{inv:5}.
Therefore, we need to protect both the Stream Table and the stage-2 translation tables as shown in Figure~\ref{fig:smmu-state}. 
Further, we need to protect the SMMU's memory-mapped configuration registers as they can be used to compromise the security of the system (e.g., turn off the SMMU, turn off stage-2 translations).
In \codename, the monitor checks all updates to the SMMU. It introduces \codename interface, which the hypervisor has to use to perform its operations on the SMMU. 
Specifically, the monitor uses the list of realm devices that it maintains and only allows the hypervisor to change stage-2 translations of non-realm devices. 
For updates to the SMMU's configuration, the monitor maintains an allow-list of non-security critical fields and only allows the hypervisor to update them. 
To enforce~\ref{inv:5}, for every update to the SMMU's stage-2 translation tables for realm devices, %
the RMM ensures that: (a) the PA is not mapped in any other device's stage-2 translation tables (~\ref{inv:5}) by checking a reverse mapping;\footnote{For the reverse mapping, the RMM can maintain reverse page tables or an array indexed by PA. Currently, RMM does the latter.}  
(b) the IPA is not already mapped in the stage-2 table for this StreamID.
Further, the RMM ensures~\ref{inv:3} by checking that for every update to the stage-2 table of the SMMU, the corresponding IPA to PA mapping exists in the stage-2 tables of the owner realm VM. 

To ensure~\ref{inv:2} during the device attach process discussed in Section~\ref{ssec:device-initialisation}, \codename monitor also maintains a list of all realm device StreamIDs as shown in Figure~\ref{fig:smmu-state}. 
While attaching a device to a realm VM, the monitor locks the Stream Table before processing the request from the RMM (as explained in Section~\ref{ssec:device-initialisation}). It then checks that the StreamID for the device is not already assigned (\ref{inv:2}) in its list and creates a new STE in the Stream Table. If the StreamID is already assigned, it aborts the realm creation process.

The SMMU performs root memory accesses for stage-2 translations (step 5 in Figure~\ref{fig:smmu-state}).
Since the SMMU is trusted hardware, it can originate accesses in root mode to manage the stream and translation tables, irrespective of the world of the transaction itself. 
Independently, the device access is also subject to a GPC check, based on what world the device belongs to. 
Hence, the SMMU can perform stage-2 translations for all devices correctly while ensuring that the world isolation for memory accesses from devices is still enforced.

\paragraph{Security Decisions.}
The RME-DA allows the RMM to directly manage the realm streams. 
In \codename, we make a conscious choice to move all the SMMU operations to the monitor, such that neither the untrusted hypervisor nor the trusted RMM can directly change it. Our choice is based on two main reasons: 
(a) the SMMU has several global configuration values, which if set incorrectly by the RMM or intentionally tampered by the hypervisor, can break the security guarantees;
(b) If we allow the hypervisor and the RMM to simultaneously update the SMMU, it will open several attack vectors such as Iago attacks~\cite{checkoway2013iago}, race-conditions~\cite{controlled-data-race}, and re-entrancy attacks. Defending against these attacks is challenging and requires careful handling of legacy hypervisor code as well as formal treatment of the resulting design~\cite{via, tfx}. 
Instead, we choose to implement a centralized SMMU interface in the monitor which enforces our security invariants and serves as a single vantage point to enforce checks that disallow the hypervisor from enabling unsafe features such as debugging. 
Case in point, RME-DA allows the device to use a performance feature called Address Translation Services (ATS) wherein the device can bypass the SMMU translations and directly access host memory using real physical addresses.  
Our monitor explicitly disallows this and several other features. Specifically, \codename maintains a limited allow-list of safe SMMU features, and rejects any requested settings that violate it.

\subsection{MMIO and DMA}
\label{ssec:mmio-dma}
In \codename, all memory-mapped regions for a device are transitioned to realm world and mapped to the realm VM as part of the device attach process as discussed in Section~\ref{ssec:device-initialisation}. 
The cores directly issue all the reads and writes to the memory-mapped regions without the involvement of the RMM or the monitor. 

Applications use DMA to transfer large amounts of data between host and device memory. In \codename, this has to be mediated via \nss. 
The realm VM's OS builds a scatter-gather list of the memory to be marked as \nss memory before initiating the DMA operation. 
To enable secure DMA and enforce~\ref{inv:3} and~\ref{inv:5}, \codename introduces a new RMM interface call for the realm VM's OS. 
Once this scatter-gather list is setup, the OS can invoke this interface to transition the DMA memory to the \nss. 
For this, the OS sends a list of IPA and the size of each region along with the device id to the RMM. 
The RMM services this request by first translating the IPA to PA using its stage-2 translation tables for the realm VM. 
This process ensures~\ref{inv:3} and~\ref{inv:5} because the same IPA to PA mapping is used to update the SMMU's stage-2 translation table.
Note that, all memory transitioned to protected memory this way already belongs to the realm VM and abides by~\ref{inv:4}. But, allowing the device to access this memory requires explicit steps.
To transition the memory to \nss, the RMM invokes the monitor which updates the SMMU's stage-2 tables as discussed in Section~\ref{ssec:secure-smmu-setup}.
After this, the device can access the \nss via DMA.
This approach only changes the Linux Kernel. It requires no changes to the application (on both the processor and the device), the device driver, the runtimes, and accelerator logic. 

\paragraph{Transparent Encryption \& Decryption.}
In CCA, the main memory connection to the bus on the host includes a memory protection engine (MPE) that encrypts data before storing and decrypts it on the way to the bus. 
It relies on the Memory Encryption Contexts (MEC) extension which is a part of Arm CCA~\cite{arm-mec}. 
With MEC, CCA allows each realm VM to be associated with a unique key. 
Then the MPE ensures that all data from a realm VM, say with ID $R_1$, is encrypted with its own key (say $K_{R1}$). 
\codename leverages the MPE and MEC to protect data stored in host DRAM. 
For PCIe communication, \codename uses PCIe IDE support in the host and accelerator root ports to protect the data while it transits the PCIe links.
\begin{figure}[]
\centering
\includegraphics[scale=0.675]{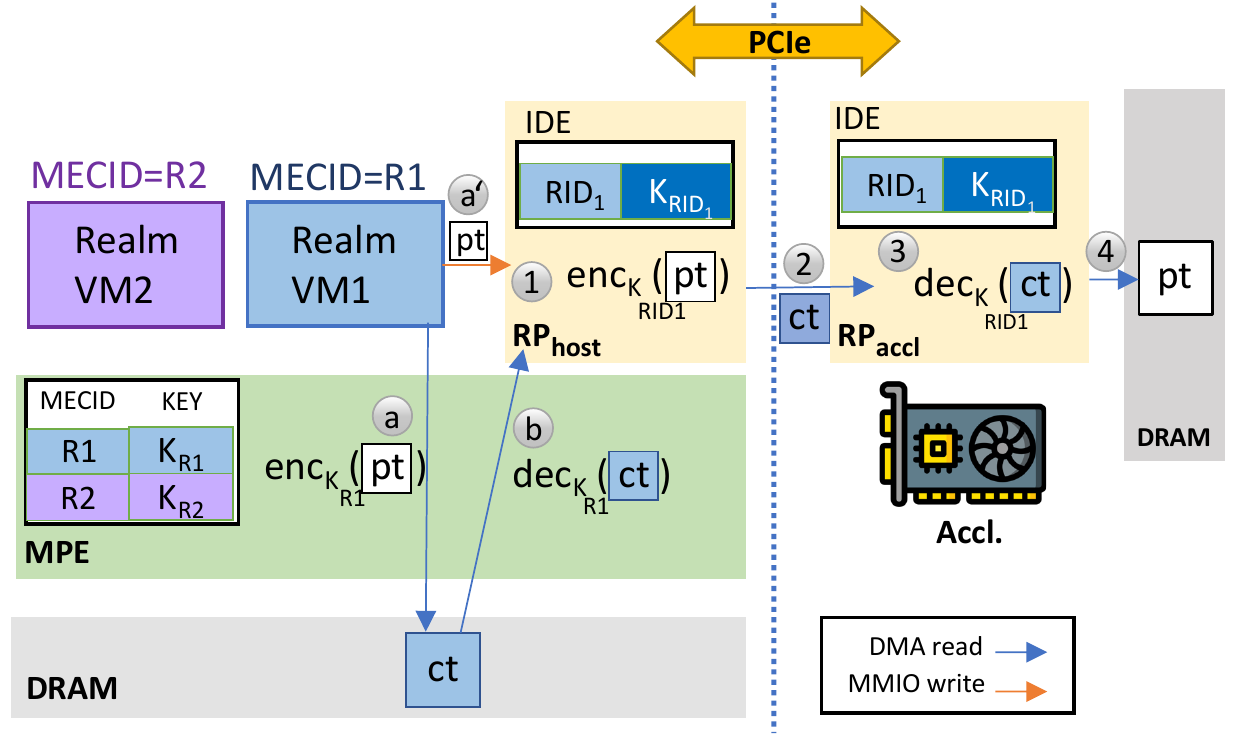}
\spacesave
\caption{MMIO and DMA in \codename. 
IDE keys in  Root Ports (RPs) are setup by the monitor on device attach. 
Blue arrows show DMA read and Orange arrows show MMIO write.         
} 
\label{fig:mmio-dma}
\end{figure}

Concretely, the following steps are performed for DMA as summarized in Figure~\ref{fig:mmio-dma}.
When a realm VM writes plaintext data $\tt{pt}$ to memory, the MPE encrypts $\tt{pt}$ to form ciphertext $\tt{ct}$ (step a). 
Further, when the accelerator initiates a DMA read, the MPE decrypts $\tt{ct}$ and sends it to the RP\textsubscript{host} (step b).
Next, RP\textsubscript{host} uses the RID\textsubscript{1} to encrypt $\tt{pt}$ to form $\tt{ct}$ (step 1). 
The $\tt{ct}$ is then transmitted over PCIe to the accelerator (step 2), to reach RP\textsubscript{accl}. It uses the RID\textsubscript{1} to decrypt $\tt{ct}$ (step 3), 
and finally stores $\tt{pt}$ in the accelerator's DRAM (step 4).
Similarly, for MMIO (step a'), when a realm VM writes ($\tt{pt}$) to a memory-mapped region of the accelerator the write is forwarded to the RP\textsubscript{host} by the MMU. Then, $\tt{pt}$ is transferred to the accelerator (steps $1-4$).
Note that, $\tt{pt}$ in step 1 is safe from a physical attacker as it does not leave the Arm SoC before being encrypted by the RP\textsubscript{host}. Similarly, step 4 writes $\tt{pt}$ to the accelerator's tightly integrated DRAM which is safe from a physical attacker~\cite{volos2018graviton}.

\spacesave
\section{Security Analysis}
\spacesave

\codename's invariants to setup device identities and memory protection achieve secure composition of CPU and device TEEs. 
Any attacks that remain would either break guarantees provided by the CPU TEE (i.e., CCA) or the device TEE itself.

\paragraph{Untrusted Hypervisor.}
It can compromise the MPE and IDE encryption keys. 
However, as explained in Section~\ref{ssec:device-initialisation}, these keys are protected in the root mode and inaccessible to the hypervisor. 
The hypervisor can emulate PCIe devices and trick \codename into attaching them to realm VMs. Such attempts would be detected through attestation according to~\ref{inv:1}. 
Specifically, an emulated device cannot produce an authentic attestation report and would be detected by the remote verifier. 
Further, the hypervisor cannot emulate a device after attestation because the configuration space is locked by the monitor and the IDE keys are programmed before gathering the attestation report. 
During device attach, the hypervisor can allocate incorrect BAR region sizes, but it will be detected during attestation. 
Further, the hypervisor can allocate memory for the BAR regions in the normal world, instead of realm world. 
However, the RMM ensures that all memory-mapped regions are transitioned to realm world and assigned to the realm VM to ensure~\ref{inv:4}. 

The hypervisor can change the PCIe configuration space of a realm device to setup memory-mapped regions in the normal world. 
Similarly, it can provision DMA memory for the realm devices in the normal world. 
Such memory mappings to the normal world memory would compromise security guarantees that \codename accords to computation on the accelerator. 
These attacks are thwarted by \codename by ensuring that all device memory and configurations are mapped solely to the realm VM that owns the device according to~\ref{inv:2} and~\ref{inv:5}. 
The hypervisor can register a malicious realm device with the same identity (RequesterID) as a victim realm device. 
To enforce~\ref{inv:1} and~\ref{inv:2}, the monitor always checks that a device is not already linked to a realm VM before assigning it.
Therefore, such attacks from the hypervisor will be detected and stopped.
The RMM exposes an interface to the hypervisor to allow it to transition realm memory back to normal world. 
While a device is performing DMA to a region of memory, the hypervisor can invoke the RMM to transfer this memory back to normal world. 
If this succeeds, the device would write sensitive data to unprotected normal world memory. 
However, CCA specification does not allow the hypervisor to transition realm memory that is still attached and in use by the realm VM thus stopping such attacks~\cite{arm-rmm-spec}. 

\paragraph{SMMU Configuration.}
The untrusted hypervisor can: (a) reconfigure the SMMU to bypass stage-2 translations,   
and (b) rewrite the stage-2 translation tables to allow attacker-controlled devices to access the protected region. 
The monitor prevents these attacks by storing all SMMU data structures (Figure~\ref{fig:smmu-state}) in root memory to enforce~\ref{inv:2}. 
Without direct access to the SMMU, the hypervisor can still setup overlapping mappings with a malicious realm device (e.g., map the same PA to the victim and malicious device during device attach). 
This attack is thwarted by the RMM which enforces~\ref{inv:5}. 
Further, the hypervisor can set up SMMU mappings to allow a normal world device to access protected memory. 
When the normal world device originates such an access to realm memory, the GPC on the SMMU will deny it. 

\paragraph{Malicious Co-tenants.}
A malicious realm VM can attempt to set up overlapping DMA memory regions with a victim's device to access its memory. 
As per~\ref{inv:4}, the RMM checks its stage-2 translation tables to find that the malicious realm VM is requesting DMA mappings for PAs it does not own, and denies such mapping requests. 

\paragraph{Other Malicious Devices.}
They can originate PCIe packets with the identity (RequesterID) of a victim device attached to a realm VM. 
As per ~\ref{inv:1}, the monitor ties the identity (RequesterID) of the realm device to the IDE key used for encryption.
Therefore, any packet that the malicious device originates with the victim's identity will fail to decrypt successfully at the root port.
Malicious co-tenant realm devices can originate PCIe traffic to access PAs that are mapped to victim devices. 
\codename uses the SMMU's stage-2 translation tables to enforce~\ref{inv:5} that ensures that these accesses are not possible. 
In particular, the stage-2 translation tables ensure that there is no valid mapping for the malicious device to access the victim device's protected realm region. 
Further, the PCIe network is not trusted and can have untrusted devices on path that can intercept the communication. 
However, all traffic between the host and the device is encrypted using IDE and cannot be compromised by such on-path devices.

\paragraph{Physical Attacker.}
It can read, inject, modify, record, and replay PCIe packets on the untrusted PCIe network (e.g., PCIe switches, interposers). 
\codename uses PCIe IDE to ensure~\ref{inv:2} and encrypts all its PCIe traffic to thwart such attacks. 
Further, the attacker can read data from the host DRAM  (e.g., cold boot attacks) which are stopped by CCA's MPE based encryption. 
A physical attacker can plug in compromised devices (e.g., debugging features enabled, old firmware). Such devices, if attached to a realm VM, can compromise its security. 
\codename relies on attestation to ensure that the device attached to a realm VM is properly configured and expects the remote verifier to check for misconfigured devices. 
Further, a physical attacker can plug in or route the PCIe packets to a malicious device after the attestation checks pass.
However, such a device will not have the IDE keys to encrypt/decrypt communications with the host realm VM.

\spacesave \lessspacesave
\section{Implementation}
\label{sec:implementation}

\begin{table}
    \centering
    \setlength{\extrarowheight}{0pt}
    \addtolength{\extrarowheight}{\aboverulesep}
    \addtolength{\extrarowheight}{\belowrulesep}
    \setlength{\aboverulesep}{0pt}
    \setlength{\belowrulesep}{0pt}
    \caption{New interfaces and updates introduced by \codename.}
    \label{tab:interfaces}
    \footnotesize
    \resizebox{\columnwidth}{!}{%
\begin{tabular}{p{0.35\columnwidth}p{0.08\columnwidth}p{0.6\columnwidth}}
\toprule
API                 & Status  & Description                                        \\ \midrule
\rmidatacreate   & changed & add data from normal world to realm memory. \codename adds $\tt{attach\_dev}$ flag. \\ \hline
\rsidelegateprotmem & new & delegate realm memory to protected memory. calls \smcdelegateprotmem.      \\ \hline
\smcdeviceattach & new     & attach and detach a device from realm.              \\ \hline
\smcdelegateprotmem & new & delegate realm memory to protected memory. add stage-2 translation for the SMMU. \\ \bottomrule
\end{tabular}%
}
    \end{table}
    
We prototype \codename on ARM Fixed Virtual Platform (FVP) with RME-support and Cortex-A53 board with no RME. 
\subsection{Putting it together}
\label{sec:putting-it-together}
\paragraph{Existing APIs.}
\codename does not make any changes to CCA or Arm architecture. Instead, it extends the monitor and RMM software interface to realize its mechanisms (see summary in Table~\ref{tab:interfaces}). 
We first explain the existing mechanism for creating realm VMs with CCA.
As per Arm's specification, 
the hypervisor creates the realm VM and then transitions it to the realm world.
When the RMM receives such a new realm creation request from the hypervisor, the RMM creates VMID, a unique ID per realm VM and uses it in all its internal data structures to track contexts that refer to this VM.
Then the hypervisor transfers the VM to the realm world using two RMI calls: \rmigranuledelegate moves memory from normal to realm world; \rmidatacreate populates data and assigns memory to the realm VM.
For the data, the RMM copies contents from a source granule ($\tt{src}$) to the destination granule ($\tt{dst}$).
Note that, $\tt{src}$ and $\tt{dst}$ are sent as arguments to the RMI call. 
Further, the RMM maps the destination granule to the $\tt{ipa}$ from the arguments of the call in its stage-2 translation tables, after checking that it is valid. 
A mapping is valid if 
(a) the PA, in this case the physical address of the $\tt{dst}$, is not mapped to any other realm VM; and (b) the $\tt{ipa}$ is not already mapped in RMM's stage-2 translation tables.

\paragraph{FVP.}
Arm Fixed Virtual Platforms (FVP) are simulations of Arm systems, including processors, memory, and peripherals with annotations to run at speeds of real hardware. 
We prototype \codename on Arm FVP\_Base\_RevC-2xAEMvA with RME-support.
We considered other open-source alternatives such as Samsung Islet~\cite{samsung-islet}, Huawei QEMU RME-support extension~\cite{huawei-qemu}, Twinvisor~\cite{twinvisor}, and virtCCA~\cite{virtcca}. However, we decided to use the FVP with the Trusted Firmware-a  (TF-A) and TF-RMM~\cite{arm-tfa, arm-rmm} because they are officially supported by Arm and best represent Arm's hardware and software stack. 

\paragraph{\tfa.}  
It is an open-source software for the EL3 monitor~\cite{arm-tfa}. 
We augment it to protect and manage stage-2 translation tables and configuration registers of the SMMU.
We implement two new SMCs in the \tfa (See Table~\ref{tab:interfaces}). 
The SMC \smcdeviceattach exposes an interface for the RMM to attach and detach a device to a realm VM. 
The FVP models PCIe devices connected to an SMMU. While we cannot use these modelled devices to generate traffic, we can use them in our implementation to measure the overheads of attaching a device. 
To this end, \codename probes the PCIe bus and reads the configuration space of the device while attaching it. 
To detach a device, we probe the PCIe bus to emulate a device reset.
With this approach, we cannot model device-side attestation, IDE key programming, and MPE.
We add \smcdelegateprotmem to enable the RMM to delegate device memory to the protected region and setup stage-2 translations (c.f. Section~\ref{ssec:secure-smmu-setup}). 
Further, we implement an \codename interface to securely allow the hypervisor to manage the SMMU configurations for the normal world.

\paragraph{RMM.} We extend the open-source TF-RMM implementation with a new \rsidelegateprotmem for the realm VM to protect device memory during runtime, as discussed in Section~\ref{ssec:mmio-dma}. 
We also setup stage-2 translations for the SMMU as explained in Section~\ref{ssec:secure-smmu-setup}. 
To attach the accelerator to the realm VM, we add a new flag called $\tt{attach\_dev}$ to \rmidatacreate.
The hypervisor uses the \rmidatacreate with the $\tt{attach\_dev}$ flag to indicate that the data in this RMI call is used to attach a device to the realm VM.
On receiving this call, the RMM performs the checks discussed in Section~\ref{ssec:device-initialisation} and invokes \smcdeviceattach to attach a device to the realm VM. 
Further, we extend the realm VM destruction process to call  \smcdeviceattach to detach a device, if present. 

\paragraph{Linux.} In the normal world, we boot Linux in EL2 with KVM patches that add support for RME and the creation of realm VMs.  
We change Linux's Arm SMMU driver to make SMC calls to configure the SMMU and its stage-2 translation tables by adding $636$\,LoC. 
We use $\tt{kvmtool}$, a lightweight tool for hosting KVM guests, from Arm to build and deploy realm VMs from the normal world. 
We augment $\tt{kvmtool}$ to be able to invoke \rmidatacreate with the \attachdev flag. 
For the realm VMs, we build Linux with patches to enable communication with the RMM.
We extend the Linux kernel and add a kernel module to the realm VM to call \rsidelegateprotmem as explained in Section~\ref{ssec:mmio-dma}.

\begin{figure}[]
    \centering
    \includegraphics[scale=0.5]{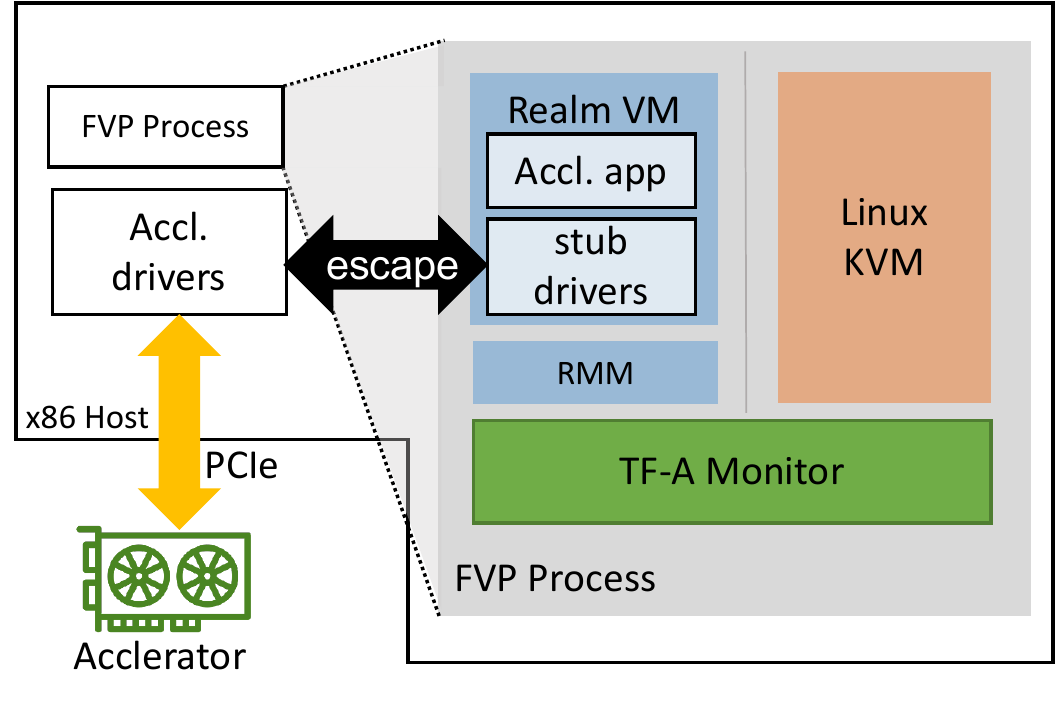} 
    \spacesave
    \caption{Experimental Setup. The full device drivers run on the x86 host. In the realm VM we run stub drivers that enable the FVP-escape mechanism. The x86 host setup is only used for evaluation and does not impact the security of \codename.}
    \label{fig:fvp-setup}
\end{figure}

\paragraph{Device Communication.} The Arm FVP does not provide functional interfaces to connect PCIe devices.
Our host machine is an x86 workstation connected to PCIe devices. It runs the FVP user process, as shown in Figure~\ref{fig:fvp-setup}. 
We have to allow the realm VM emulated in the FVP process to communicate with the real PCIe devices connected to our host machine. To enable this, we implement an FVP escape mechanism. 
We modify the Linux kernel on our x86 host machine to enable devices to directly access the FVP user process's memory. 
To allow applications running in the realm VM on the FVP to communicate with the devices, we implement stub drivers. 
On the x86 host, we register page-fault handlers to memory regions in the FVPs address space. 
When the stub drivers access these memory regions, the fault handler invokes the device drivers on the x86 host to perform the operations. 
To implement these stub drivers, we utilize an open-source device stack. For the GPU, we use Gdev~\cite{gdev} as a CUDA runtime and libdrm with Nouveau from the v5.10 Linux kernel as the driver on our host machine. 
We cross-compile Gdev for ARM using Arm GNU Toolchain. 
We run our experiments on a GeForce GTX $460$ SE GPU that is compatible with Gdev and Nouveau. 
For the FPGA, we use a Xilinx Virtex Ultrascale+ VCU118 FPGA. 
We use Xilinx's XDMA driver, which we create stubs for and cross-compile for ARM.
 
\paragraph{GPU and FPGA.}
For the baseline encrypted mode, we augment the benchmark applications on the GPU with encryption/decryption kernels that use AES-CTR mode encryption with $256$-bit block size~\footnote{For simplicity, we use AES-CTR mode instead of AES-GCM.}.
Similarly, for the FPGA, we use standard implementations of AES-CTR mode from Vitis libraries to add encryption to the user logic executing on the FPGA.
Because these Vitis Libraries are not natively supported for our FPGA, we port them using Vitis HLS and Vivado to run on our board. 
To measure the costs of encryption for GPU apps on the host, we cross-compile libopenssl $1.1.1q$ to Arm. The FPGA applications are written in Python so we cross-compile Python as well as libcrypto to run on the FVP. 

\paragraph{Optimization.}
In \codename, the realm VM requests stage-2 translation table mappings for DMA during runtime. 
\codenameopt optimises this by extending the realm creation explained for \codename in Section~\ref{ssec:device-initialisation} to pre-setup SMMU stage-2 translation tables. 
Specifically, if a device is attached to the realm VM, \codenameopt adds SMMU stage-2 translation table entries for all realm VM memory before starting the realm VM. 
Further, during the runtime of the realm VM, \codenameopt ensures that any new memory added to the realm VM (via RMI calls) is also mapped in the SMMU's stage-2 translation tables. 
This approach ensures that the device has access to the entire realm VM memory eliminating the necessity of invoking \rsidelegateprotmem for each DMA. 
The trade-off is that this optimization increases the time for realm VM creation and extends realm VM memory as well as the size of the stage-2 translation tables stored in the root. 
\subsection{Performance Evaluation Prototype} 
\paraspacesave
We use the Zynq UltraScale+ MPSoC ZCU102 with an Arm Cortex-A53 $64$-bit quad-core processor to estimate  \codename performance.
We run our modified \tfa with SMMU and GPT changes in EL3. 
Because the processors do not support GPC, we estimate the overhead of programming the GPT registers by using other idle EL3 registers (i.e., AFSR0\_EL3 and AFSR1\_EL3).
The processor does not have RME support. 
Therefore, to emulate the realm world, we create a Realm context in the normal world. 
We run a modified RMM in the EL2 of the Realm context. 
To create this context, we use Xilinx PetaLinux tools that enable us to build a custom \tfa that also boots the modified RMM. 
We adapt the \tfa and RMM we run on the FVP to boot on the board. 
This requires us to replace RME specific instructions (e.g., to invalidate GPT TLBs) and adapt the translation tables to be compatible with Armv8-A.  
The RMM implementation requires that it is run in the realm world. 
In our Realm context, we ensure the right security state by setting the SCR\_EL3 register.
Xilinx Petalinux tools does not allow us to load custom stage-1 and stage-2 bootloaders. Therefore we move RME-specific initialization (e.g., GPT setup) to bl3.
Since the processors do not have RME, our measurements do not reflect true GPC cost.

\spacesave
\section{Evaluation}
\label{sec:eval}
\paraspacesave
\paragraph{Compatibility.}
We build a VM with $\tt{kvmtool}$ that includes Arm's changes to support CCA.
The VM uses a patched Linux kernel $6.2$ that supports CCA both as host and guest OS.
The only change that \codename requires is adding an \codename-helper kernel module comprising $320$\,\loc to the realm VM.
We do not change the applications, accelerator drivers, or runtimes.

\paragraph{TCB.}
Our baseline setup is a confidential VM using an accelerator device with bounce buffer design~\cite{nvidia-azure}. 
Its TCB includes the monitor, the RMM, and the guest kernel (totals $25.9$M\,\loc) as well as the device runtime, the driver, and the software implementation for memory encryption (totals $203$K\loc and $6.5$K\loc for GPU and FPGA respectively).\footnote{For simplicity, we leave out the size of software encryption implementation (e.g., ~$835$\,K\loc) of OpenSSL, from the baseline TCB.}
Our modifications to the monitor $1588$\,\loc, the RMM $382$\,\loc, and the guest kernel $1734$\,\loc are minimal. 
Overall, \codename adds $3704$\,\loc to the TCB, totalling $26.1$M\loc and $25.9$K\loc for GPU and FPGA respectively.

\subsection{Experimental Setup}
\paraspacesave
We run our experiments on an x86 host with a $32$-core dual-socket Intel Xeon Gold $6346$ CPUs and $378$ GB RAM. 
\label{sec:eval-setup}

\paragraph{Benchmark Selection.}
\begin{table}[t]
\centering
\caption{Overview of GPU Benchmarks. Task refers to the number of CUDA kernels launched on the GPU. T Size: Transfer size in MB. P Size: Problem Size in points}
\tableappendixspacing

\label{tab:gpu-benchmarks}
\resizebox{\columnwidth}{!}{%
\begin{tabular}{@{}llrrr@{}}
\toprule
App & Domain & Tasks & T Size & P Size \\ \midrule
nn & Dense linear algebra & 1 & 1 & 42764  \\
gaussian & Dense linear algebra & 3148 & 38 &  1575 $\times$ 1575  \\
needle & Dynamic programming & 229 & 39 &  1840  \\
pathfinder & Dynamic programming & 5 & 20 & 50000 $\times$  100  \\
bfs & Graph traversal & 2 & 3 &  1840  \\
srad\_v1 & Structured grid & 102 & 2 &  502 $\times$  458  \\
srad\_v2 & Structured grid & 4 & 64 &  2048 $\times$  2048  \\
hotspot & Structured grid &  5 & 3 &  512 $\times$  512 \\
backprop & Unstructured grid & 2 & 71 &  262144 $\times$  16 $\times$  1  \\ 
\bottomrule

\end{tabular}%
    }
\end{table}
\begin{table}[t]
\centering
\caption{Overview of FPGA Benchmarks }
\tableappendixspacing
\label{tab:fpga-benchmarks}
\resizebox{0.9\columnwidth}{!}{%
\begin{tabular}{@{}llrrr@{}}
\toprule
App & Domain & T Size  & P Size \\ \midrule
matmul5 & Matrix Multiplication & 300 B &   5 $\times$ 5 \\
matmul10 & Matrix Multiplication & 1200 B &   10 $\times $ 10 \\
svd32 & Singular Value Decomposition & 32 KB &  32 $\times $ 32 \\
svd64 & Singular Value Decomposition & 128 KB &  64 $\times $ 64 \\
\bottomrule
\end{tabular}%
}
\end{table}
We select a standard set of GPU benchmarks from the Rodinia testsuite~\cite{rodinia}. Instead of the latest Rodinia release from gdev-bench~\cite{gdev-rodinia}, we use version $2.1$ since it uses CUDA Driver APIs and is compatible with Gdev.
Of the $11$ available benchmarks, we successfully test $9$. We exclude $\tt{lud}$ and $\tt{heartwall}$ because they crash even on the unmodified system, probably because they are incompatible with our GPU. 
Nevertheless, as shown in Table~\ref{tab:gpu-benchmarks}, 
the benchmarks cover a wide range of domains, data workloads, and sizes of applications.
We fail to identify a standard benchmark that is used for FPGAs. 
We surveyed the hand-coded benchmarks from prior works~\cite{zhao2021shef}.
Some of them are not supported on our FPGA (e.g., DNNWeaver runs on KCU1500). 
Our choice of FPGA was based on what is available on cloud FPGAs (e.g., Amazon F1).
However, custom-coded apps for the cloud do not run on local FPGAs because they expect Amazon shell on AWS.
Thus, we hand code Matrix Multiplication and Singular Value Decomposition (see Table~\ref{tab:fpga-benchmarks}) using existing algorithms from Vitis Libraries and Vitis HLS.

\begin{figure}[t]
    \includegraphics[scale=0.55]{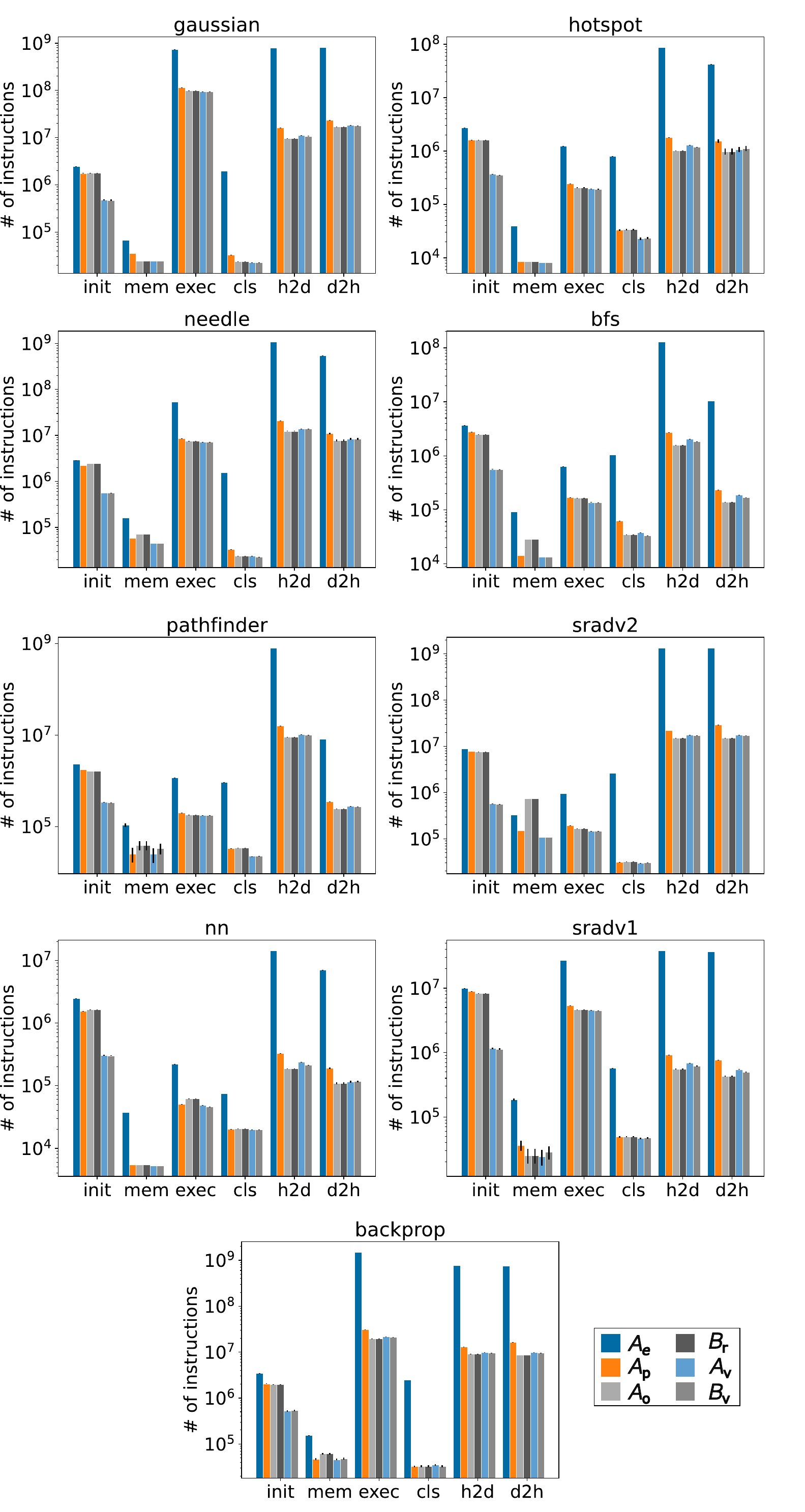}
    \caption{Overheads for GPU benchmarks.}
    \label{fig:gpu-perf}
    \vspace{-4.5em}
\end{figure}

\paragraph{Benchmark Configurations.}
Our vanilla setup \bvan which comprises an unmodified monitor, RMM, and hypervisor, 
measures the benchmark performance on a VM in the normal world.
To measure the effect of \codename changes on normal world operations (\avan), we execute the benchmarks in the normal world VM but with a modified monitor, RMM, and hypervisor. 
For measuring \codename in encrypted mode (\aenc) we launch the same setup as above but run the benchmarks in a realm VM equipped with software encryption. 
Then, we enable \codename protected mode (\aprot) and measure the benchmark performance with a modified monitor, RMM, hypervisor, guest Linux kernel hooks, and SMMU. 
Comparing \aenc and \aprot, we can quantify the improvements due to \codename protection mode. 
Note that \aprot and \aenc include the pure cost of CCA's protection. 
To decouple the \codename cost from baseline CCA cost, we setup a configuration where we launch the realm with unmodified software (\brl).
Thus, we can compare the performance of \codename (\aprot and \aenc) with normal world execution (\bvan and \avan) by deducting the costs of CCA (\brl).
Further, we do the optimization (\aopt) (c.f. Section~\ref{sec:putting-it-together}) and compare it to \aprot. 

\paragraph{Arm FVP.} Our FVP base model, the proprietary version publicly offered by Arm, emulates $8$ Arm cores with features from v$8.2$ to v$9$. 
We use the Model Trace Interface (MTI) in the FVP to count the number of instructions and events. 
This latest version of the FVP (FVP\_Base\_RevC-2xAEMvA) has support for RME. We thus use this model to emulate Arm CCA and realm VMs. 
The FVP simulates a limited implementation of the PCIe standard and does not support real device connections over PCIe. 
So the physical device (e.g., a real GPU) and FVP cannot access each other's memory to send and receive data. 
We address this in two steps: 
First, we run the FVP user process on a host machine that is connected to a PCIe device. We then add an escape mechanism on the host machine such that the device can access (parts of) the FVP process memory as shown in Figure~\ref{fig:fvp-setup}. 
Specifically, when the VM inside the FVP allocates device memory, our escape mechanism presents this as a host\\ 
\\ 
\\  
memory to the PCIe device. 
This allows us to emulate direct device accesses from the 
FVP---when the VM makes DMA or MMIO operations, they are visible to the PCIe device and vice versa. 
Our SMMU GPT and stage-2 translations are setup to protect this memory.
However, the FVP escape mechanism bypasses the SMMU in the
FVP. 
Our second step addresses this issue. 
The FVP's CPU cores are connected to a SMMU engine that is driven by a stub. The stub can generate test traffic.
So we record PCIe transactions on a real host-device setup and replay it to generate traffic plugged to the SMMU via the stub. 
This ensures that we trigger the SMMU for all device-related operations. 

With this, one problem still remains. 
In the FVP, the SMMU is not RME-DA compliant.
So, its GPC does not allow access to realm memory where \codename creates the protected region. 
Because we cannot change the FVP, we remedy this issue by using two GPTs---one for the cores (\coregpt) and one for the SMMU (\devgpt)---to emulate the RME-DA functionality. 
The first \coregpt marks the protected memory as realm world for all cores. 
This way, only cores executing in realm mode (realm VMs and RMM) can access this
region. We use this GPT for all CPU cores.
On the other hand, the accelerator cannot access the shared memory if the GPT
marks it as realm world. To address this, we create a second GPT (\devgpt) where
the shared memory is marked as normal world.
We then assign this GPT to the SMMU, thus enabling it to access the
shared region. 
Whenever the monitor makes changes to the core GPT, we update the
SMMU's GPT to ensure that the accelerator does not have access to any
protected or realm memory that the core cannot access. 
We change the \tfa stage-2 and stage-3 boot loaders to create the additional \devgpt. 
Our accelerators and host are not \pciefive enabled, so we fall back to PCIe-4 and do not include one-time IDE setup cost. 

\paragraph{Arm Board.}
The FVP is instruction-accurate, i.e., it reports the exact number of instructions that an Arm core will execute to perform operations. 
However, it is not cycle-accurate, i.e., we cannot use it to measure the wall-clock performance of an out-of-order core operating at a certain frequency. 
So while our instruction count measurements on the FVP are a proxy metric to estimate performance, they cannot be used to project real-time performance.
Since there are no public performance reports for Armv9 cores with RME enabled for CCA, we cannot reliably predict the performance of any CCA workloads. 
As a best effort, we measure the costs of specific operations on a real Arm board with Cortex-A53 cores.%
We optimistically assume that Arm cores for CCA will perform close to traditional virtualization and calculate the projected performance of our workloads, while accounting for operational frequencies and average instructions executed per cycle.
We do not claim that these performance numbers from the FVP and the board reflect real Arm CCA CPUs.

\subsection{Lifecycle Costs \& \codename Breakdown}
\label{sec:eval-stats}
    \begin{figure}[]
        \includegraphics[scale=0.55]{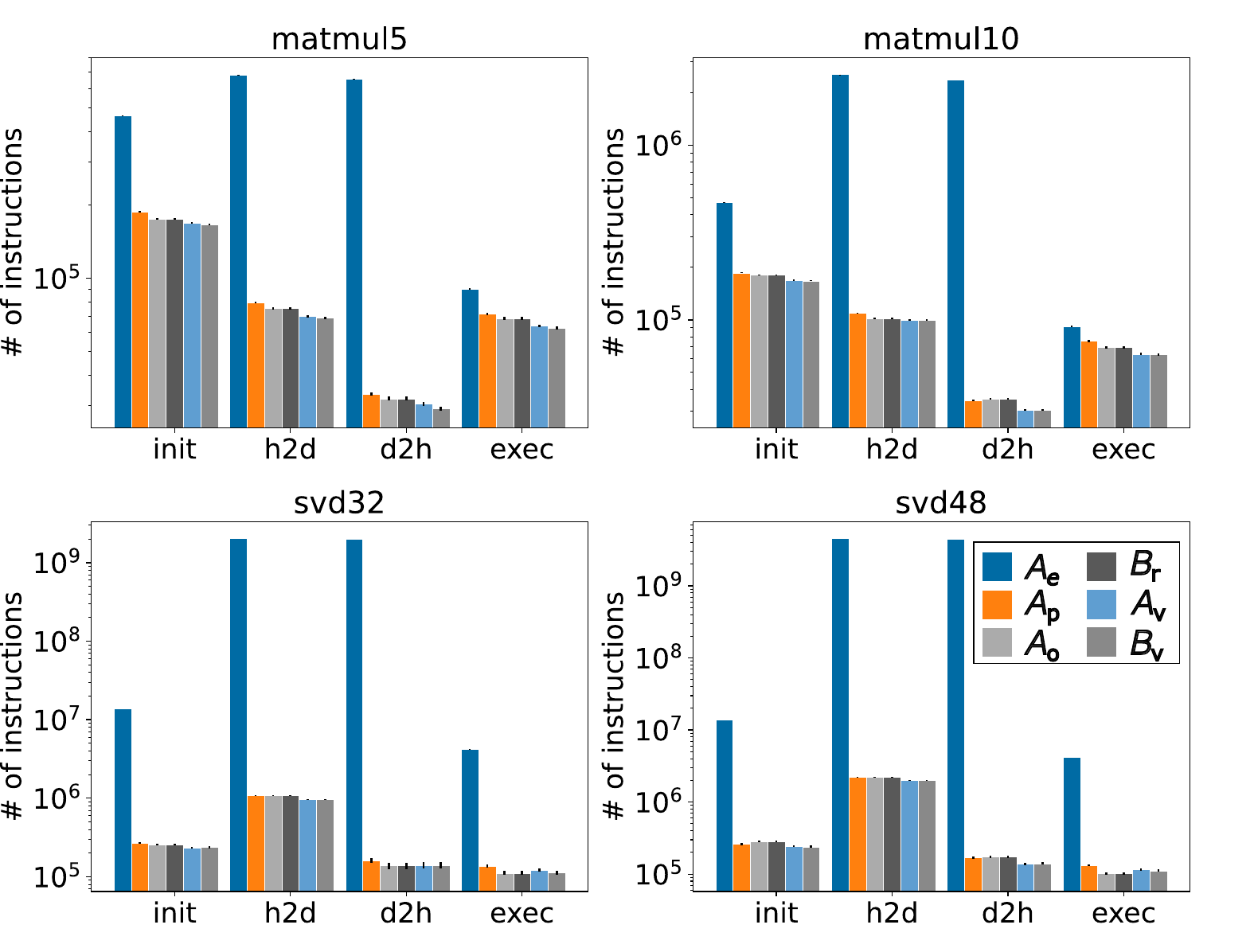}
        \caption{Overheads for FPGA benchmarks. We do not measure mem and cls stages for these benchmarks as they do not perform dynamic memory allocations. }
        \label{fig:fpga-perf}
    \end{figure}

\paragraph{System Boot and Realm Creation.} During system boot, \codename adds a minimal overhead of $11.42\%$ to delegate the SMMU's data structures to the root world as compared to \bvan. Note that this overhead is a one-time cost.
The realm creation operation is performed once per realm VM. The costs depend on the VM size and the reserved device memory size. Most init time is spent in transitioning pages from normal to realm world. 
Compared to just creating realm VMs (\brl), \codename adds $5.34\%$ overheads due to the additional RSI and SMC calls.

\paragraph{Deletion.}
\codename does not add overhead for clean-up.
The realm memory footprint of \codename is the same as the baseline (\brl). 
For releasing the device, the only overhead is to delete the stream table entry for the SMMU ($69$ instructions).
The subsequent TLB flush in the SMMU is necessary for any device deletion, even for the baseline.
The cost of cleaning the device state such that it can be allocated to another VM is common across baselines, \codename does not add any overheads.

\begin{table*}
    \centering
    \caption{GPU Performance. Transfer size in MB. Interface calls, Context Switches \& World Switches for \aprot. \textit{RSI prot} refers to  \textit{rsi\_delegate\_prot\_mem} calls. Rl: Realm, Rt: Root, and N: Normal world. Setup \& Runtime costs: E\% percentage of the total execution for the stage, O\% overheads of \aprot to \bvan. \textit{H to D} and \textit{D to H} refer to host to device transfers and vice-versa. \aprot includes CCA protection overheads not present in \bvan. Negative overheads imply \aprot is faster than \bvan.}
    \tableappendixspacing
    
    \label{tab:gpu-bench}
        \resizebox{\textwidth}{!}{%
    
    \begin{tabular}{lrrrr|rrrr|rrr|rrrrrr|rrrrrr} 
    \toprule
     &  & \multicolumn{3}{l}{\makecell{Interface \\ calls ($10^3$)}} & \multicolumn{4}{c}{\makecell{EL switches \\ from ($10^3$)}} & \multicolumn{3}{c}{\makecell{World switches \\ from ($10^3$)}} & \multicolumn{6}{c}{Setup (\%)} & \multicolumn{6}{c}{Runtime (\%)} \\ 
    \midrule
     & \multirow{2}{*}{\begin{tabular}[c]{@{}c@{}}transfer \\ size\end{tabular}} & \multicolumn{1}{l}{\multirow{2}{*}{\# dma~}} & \multirow{2}{*}{\begin{tabular}[c]{@{}c@{}}RSI\\ prot~\end{tabular}} & \multirow{2}{*}{\begin{tabular}[c]{@{}c@{}}RMI\\ calls~\end{tabular}} & \multirow{2}{*}{EL0} & \multirow{2}{*}{EL1} & \multirow{2}{*}{EL2} & \multirow{2}{*}{EL3} & \multirow{2}{*}{Rl} & \multirow{2}{*}{Rt} & \multirow{2}{*}{N} & \multicolumn{2}{c}{init} & \multicolumn{2}{c}{memalloc} & \multicolumn{2}{c|}{close} & \multicolumn{2}{c}{exec} & \multicolumn{2}{c}{H to D} & \multicolumn{2}{c}{D to H} \\
     &  & \multicolumn{1}{l}{} &  &  &  &  &  &  &  &  &  & E\% & O\% & E\% & O\% & E\% & O\% & E\% & O\% & E\% & O\% & E\% & O\% \\ 
    \midrule
    nn & 0.5 & 0.2 & 0.1 & 0.1 & 0.2 & 0.2 & 0.5 & 0.2 & 0.1 & 0.2 & 0.1 & 72.0 & 411.2 & 0.3 & 3.8 & 1.0 & 2.4 & 2.4 & 9.7 & 15.5 & 54.2 & 8.9 & 63.1 \\
    gaussian & 37.9 & 12.9 & 8.1 & 1.1 & 38.0 & 46.8 & 19.4 & 10.4 & 9.2 & 10.4 & 1.1 & 1.1 & 273.4 & 0.1 & 43.8 & 0.1 & 46.3 & 73.5 & 21.5 & 10.3 & 49.3 & 15.0 & 30.5 \\
    needle & 38.8 & 10.2 & 5.0 & 0.3 & 3.4 & 8.6 & 11.1 & 5.7 & 4.4 & 4.0 & 2.7 & 5.1 & 295.1 & 0.1 & 28.4 & 0.1 & 45.4 & 20.1 & 21.2 & 49.1 & 52.8 & 25.5 & 31.3 \\
    pathfinder & 19.3 & 5.0 & 2.9 & 0.1 & 0.2 & 3.1 & 6.3 & 3.2 & 1.8 & 5.4 & 0.1 & 9.6 & 428.1 & 0.1 & -24.6 & 0.2 & 45.9 & 1.1 & 12.9 & 87.1 & 59.8 & 1.9 & 29.8 \\
    bfs & 3.3 & 0.9 & 0.5 & 0.1 & 0.3 & 0.7 & 1.5 & 0.8 & 0.6 & 0.7 & 0.1 & 46.5 & 395.4 & 0.2 & 5.2 & 1.0 & 88.8 & 2.8 & 22.8 & 45.5 & 45.8 & 3.9 & 39.3 \\
    srad\_v1 & 1.8 & 0.8 & 0.5 & 0.5 & 3.3 & 2.9 & 4.0 & 1.6 & 1.0 & 1.5 & 0.5 & 55.5 & 684.9 & 0.2 & 27.7 & 0.3 & 3.8 & 33.5 & 20.2 & 5.7 & 48.7 & 4.8 & 54.8 \\
    srad\_v2 & 64.0 & 16.5 & 8.4 & 5.1 & 4.9 & 12.8 & 33.6 & 18.8 & 9.3 & 18.6 & 9.3 & 13.1 & 1279.7 & 0.3 & 41.4 & 0.1 & 4.8 & 0.3 & 33.7 & 36.8 & 29.6 & 49.4 & 71.0 \\
    backprop & 71.0 & 18.3 & 9.3 & 0.4 & 0.3 & 9.7 & 19.8 & 10.1 & 5.2 & 14.7 & 0.4 & 3.2 & 280.4 & 0.1 & -3.0 & 0.1 & -0.4 & 49.8 & 47.7 & 20.7 & 36.0 & 26.1 & 71.5 \\
    \multicolumn{1}{l}{hotspot} & 3.0 & 0.8 & 0.5 & 0.1 & 0.3 & 0.8 & 1.4 & 0.7 & 0.6 & 0.7 & 0.1 & 30.6 & 353.8 & 0.2 & 3.7 & 0.6 & 41.0 & 4.7 & 26.4 & 34.7 & 51.9 & 29.2 & 36.7 \\
    \bottomrule
    \end{tabular}
    }
    \end{table*}

    \begin{table*}
    \centering
    \caption{FPGA Performance. Transfer size in KB. Interface calls, Context Switches, and World Switches for \aprot. \textit{RSI prot} refers to \textit{rsi\_delegate\_prot\_mem} calls. Setup \& Runtime costs: E\% percentage of the total execution for the stage. O\% overheads of \aprot to \bvan. \aprot includes CCA protection overheads not present in \bvan.}
    \tableappendixspacing
    
    \scriptsize
    \label{tab:fpga-bench}
    \resizebox{\textwidth}{!}{%
    \begin{tabular}{lrrrr|rrrr|rrr|rrrrrrrr} 
    \toprule
     &  & \multicolumn{3}{l}{\makecell{Interface calls ($10^0$)}} & \multicolumn{4}{c}{EL switches from ($10^0$)} & \multicolumn{3}{c}{\makecell{World switches from ($10^0$)}} & \multicolumn{8}{c}{Setup \& Runtime (\%)} \\ 
    \midrule
     & \multicolumn{1}{c}{\multirow{2}{*}{\begin{tabular}[c]{@{}c@{}}transfer \\ size \end{tabular}}} & \multirow{2}{*}{\# dma} & \multicolumn{1}{c}{\multirow{2}{*}{\begin{tabular}[c]{@{}c@{}}RSI\\ prot\end{tabular}}} & \multicolumn{1}{c|}{\multirow{2}{*}{\begin{tabular}[c]{@{}c@{}}RMI\\ calls\end{tabular}}} & \multicolumn{1}{c}{\multirow{2}{*}{\begin{tabular}[c]{@{}c@{}}EL0\end{tabular}}} & \multicolumn{1}{c}{\multirow{2}{*}{\begin{tabular}[c]{@{}c@{}}EL1\end{tabular}}} & \multicolumn{1}{c}{\multirow{2}{*}{\begin{tabular}[c]{@{}c@{}} EL2\end{tabular}}} & \multicolumn{1}{c|}{\multirow{2}{*}{\begin{tabular}[c]{@{}c@{}} EL3\end{tabular}}} & \multicolumn{1}{c}{\multirow{2}{*}{Realm}} & \multicolumn{1}{c}{\multirow{2}{*}{Root}} & \multicolumn{1}{c|}{\multirow{2}{*}{Normal}} & \multicolumn{2}{c}{init} & \multicolumn{2}{c}{H to D} & \multicolumn{2}{c}{exec} & \multicolumn{2}{c}{D to H} \\
     & \multicolumn{1}{c}{} &  & \multicolumn{1}{c}{} & \multicolumn{1}{c|}{} & \multicolumn{1}{c}{} & \multicolumn{1}{c}{} & \multicolumn{1}{c}{} & \multicolumn{1}{c|}{} & \multicolumn{1}{c}{} & \multicolumn{1}{c}{} & \multicolumn{1}{c|}{} & \multicolumn{1}{c}{E\%} & \multicolumn{1}{c}{O\%} & \multicolumn{1}{c}{E\%} & \multicolumn{1}{c}{O\%} & \multicolumn{1}{c}{E\%} & \multicolumn{1}{c}{O\%} & \multicolumn{1}{c}{E\%} & \multicolumn{1}{c}{O\%} \\ 
    \midrule
    matmul5 & 0.3 & 11.0 & 11.0 & 2.9 & 23.0 & 34.3 & 29.1 & 15.3 & 12.6 & 13.3 & 5.1 & 50.4 & 12.5 & 21.4 & 16.0 & 19.2 & 14.2 & 9.0 & 15.0 \\
    matmul10 & 1.2 & 11.0 & 11.0 & 3.1 & 23.0 & 34.6 & 30.6 & 16.9 & 9.8 & 24.5 & 10.4 & 45.9 & 11.1 & 26.9 & 8.9 & 18.7 & 19.2 & 8.6 & 14.1 \\
    svd32 & 32.0 & 27.0 & 27.0 & 12.9 & 45.0 & 79.1 & 95.7 & 52.8 & 40.5 & 54.8 & 13.9 & 16.5 & 14.0 & 65.5 & 11.8 & 8.3 & 21.7 & 9.8 & 14.6 \\
    svd48 & 128.0 & 36.0 & 36.0 & 20.9 & 67.3 & 110.7 & 135.7 & 76.7 & 49.8 & 76.2 & 27.8 & 9.5 & 9.5 & 79.6 & 9.6 & 4.8 & 18.4 & 6.2 & 20.9 \\
    \bottomrule
    \end{tabular}
    } 
    \end{table*}

\paragraph{Cost of Delegating SMMU Operations.}
Our monitor handles SMMU operations as described in Section~\ref{ssec:secure-smmu-setup}. \codename does not significantly change the memory footprint but simply moves it from normal world hypervisor to root world monitor.
Most noticeable is the monitor state to store realm
 device IDs, which is $64$ bits per realm device as shown in Figure~\ref{fig:smmu-state}.
In Figure~\ref{fig:gpu-perf} and Figure~\ref{fig:fpga-perf}, we see the effect of the SMC calls on hypervisor performance when comparing unmodified execution with the \codename changes (\bvan vs. \avan). 
The effect of adding and removing SMMU entries and stage-2 translation pages on TLB flushes is common for all configurations.

\paragraph{Interface Calls \& Context Switches.}
\codename incurs RSIs and SMCs necessary to setup and protect device memory with GPTs and the SMMU as shown in Table~\ref{tab:gpu-bench} and Table~\ref{tab:fpga-bench}. 
They lead to expensive exception-level switches. 
On average, we incur $103.8\%$ and $62.2\%$ more context switches when compared to an unmodified execution in the realm world (\brl) for GPU and FPGA respectively.
For GPU benchmarks we observe a higher number of switches for two reasons. 
First, the workloads perform several memory allocations. The execution is longer, which incurs more timer interrupts for scheduling the realm VM. 
Second, for large memory allocations, we observe a higher number of RSI calls.
We report that this is due to fragmentation---the hypervisor allocates non-contiguous physical pages, which in the worst case, can result in an RSI call per page to update the SMMU's stage-2 translation table. 
While we do not observe such extreme cases, \codename does not make any optimizations (e.g., contiguous memory allocation in the hypervisor). 
If enabled, such optimizations will reduce the number of RSIs. 
Consequently, the overhead for FPGA benchmarks is smaller because they allocate lesser memory. 
Table~\ref{tab:gpu-bench} and Table~\ref{tab:fpga-bench} summarize the interface calls and context switches for \aprot. 
For GPU benchmarks with long-running task, a majority of RMI calls are attributed to timer interrupts.
With \codename enabled, we see an increase of $7.7\%$ RMI calls compared to \brl.
\begin{table}[]
\centering

\caption{Measurements for GPT operations on Arm Board in cycles and \% overhead. GPT Setup: creates two GPTs during boot. GPT Delegate: delegate one granule. }
\tableappendixspacing

\label{tab:board-bench}
\resizebox{0.7\columnwidth}{!}{%
\begin{tabular}{@{}lrrl@{}}
\toprule
             & CCA & \codename & \ \ \% \\ \midrule
GPT Setup    & 84,672.0   & 102,247.0                & 20.8          \\
GPT Delegate & 8,516.0    & 8,574.5                  & \ \ 0.7           \\ \bottomrule
\end{tabular}%
}
\end{table}

\paragraph{Effect on Normal World Execution.}
We compare the performance of normal world execution that executes a (non-realm) VM connected to a GPU, with and without \codename changes (\bvan vs \avan). 
Figure~\ref{fig:gpu-perf} shows that \codename's enforcement of SMMU protection and additional GPT operations do not impact the normal world performance. 
We report an average overhead of $3.8\%$ and $1.9\%$ for GPU and FPGA benchmarks. Among the GPU benchmarks, we see that $\tt{hotspot}$ incurs the highest slowdown ($11.8\%$) even though it does not transfer the most amount of data as shown in Table~\ref{tab:gpu-bench}. 
We note that this might be because of how its memory is allocated.

\paragraph{Optimization.}
 \codenameopt (\aopt) achieves an average speedup of $26.05\%$ and $3.1\%$ compared to \codename (\aprot) for GPU and FPGA respectively as shown in Figure~\ref{fig:gpu-perf} and Figure~\ref{fig:fpga-perf}.

\paragraph{Costs on Arm Board.}
Frequent operations during a realm VM's execution, i.e., transitions to EL2 and EL3, are not expensive.
Transitioning from kernel to hypervisor takes $0.89$ $\mu s$, and from hypervisor to monitor takes $0.77$ $\mu s$. 
\codename requires the hypervisor to use the \codename interface to the monitor to update the SMMU data structures and update the SMMU's configuration registers. We see that performing an SMC call from the hypervisor and writing data to $1$ page costs $2.33$ $\mu s$ as compared to $1.75$ $\mu s$ when directly performed in the hypervisor. Similarly, calls to the monitor to perform memory-mapped register writes require $0.60$ $\mu s$ instead of $0.13$ $\mu s$ to write to the register directly in the hypervisor.

\paragraph{Prototype with an Additional GPT.}
To verify the functional correctness of \codename, we implement $2$ GPTs in our prototype as explained in Section~\ref{sec:putting-it-together}.
With this, the realm VM allocates and frees device memory triggering changes to the device GPT. 
As a workload, we run a bare-metal application in the realm. 
We perform additional RMI and SMC calls for attaching a device which incurs $100$ million extra instructions.
This leads to $99.15\%$ overhead for the realm VM boot. 
GPU workloads exhibit different numbers of device operations (between $11$--$9600$) that cause one GPT change per operation. Additionally, each GPT change only incurs $10$ extra instructions. 
Note that we report these measurements for completeness and the GPT changes do not affect \codename's performance reported in Section~\ref{sec:eval}.
We also measure the cost of the GPT operations on our performance prototype as shown in Table~\ref{tab:board-bench}.

\subsection{Comparison to Encrypted Mode}
\label{appx:eval-enc}
\paragraph{FVP.}
We compare the impact of \codename (\aprot) with the encrypted mode (\aenc) that uses a bounce-buffer design.
On average, \aenc requires $26.8\times$ more instructions than \aprot for GPU benchmarks. 
Backprop which transfers the most amount of data shows the highest slowdown of $47.8\times$. 
Similarly for FPGA, SVD requires $3271.7\times$ more instructions for \aenc because the FPGA apps poll for completion. These data transfers are small but frequent, incurring a high encryption overhead. 

\paragraph{Estimates on Arm Board.}
\begin{table}[]
    \centering
    \caption{Encryption cost estimates for the \aenc compared to \aprot on the board. The values reflect speedup (x) of \aprot compared to encryption measurements on the board and estimates from Graviton~\cite{volos2018graviton} for \aenc. Arm cores' frequency is $1.2$GHz.}
    \tableappendixspacing
    \label{tab:board-est}
    \resizebox{\columnwidth}{!}{%
    \begin{tabular}{lrrrrr}
    \toprule
     & nn & \multicolumn{1}{r}{gaussian} & needle & pathfinder & bfs \\ \midrule
    board (x) & 17.26 & \multicolumn{1}{r}{16.20} & 16.94 & 16.79 & 16.97 \\
    graviton (x) & 10.39 & \multicolumn{1}{r}{9.75} & 10.20 & 10.10 & 10.22 \\ \midrule
     & srad\_v1 & srad\_v2 & backprop & \multicolumn{1}{r}{hotspot} &  \\ \midrule
    board (x) & 16.45 & 19.3 & 16.93 & \multicolumn{1}{r}{16.69} &  \\
    graviton (x) & 9.90 & 32.2 & 10.19 & \multicolumn{1}{r}{10.05} &  \\ \bottomrule
    \end{tabular}%
    }
    \end{table}
Our FVP measurements use an expensive software-based cryptographic implementation in the realm VM for \aenc. 
As a best-effort estimate we compare the impact of \aenc on a real Arm board with support for hardware accelerated AES instruction set.
We assume zero costs for buffer management, SMMU setup in the \aenc case, and only account for the cost of encryption, decryption, and data copies.
We micro-benchmark a standard Arm encryption library and measure the performance for transferring a $4$KB page with AES-GCM $256$-bit block size.
This allows us to project the approximate overheads of \aenc, based on the total number of memory transfer operations required for each benchmark. 
Similarly, we estimate the performance of \aprot by calculating the cost of RSI calls invoked per memory allocation and deallocation operation during the benchmark execution. 
Despite the fast hardware encryption, we still expect a slowdown of $2$ orders of magnitude (c.f. Table~\ref{tab:board-est}). 

\paragraph{Estimates on Graviton.}
How does the performance of \codename compare to CPUs with native support for fast encryption?
To answer this question, we consider the setup from Graviton~\cite{volos2018graviton}, where the host CPU is Intel SGX with AES-NI support to perform AES-CTR mode encryption on $256$-bit blocks and SHA-3 for integrity protection. 
Based on their evaluation, we calculate the cost of host-side encryption for a $4$KB page and then apply that to our benchmarks. 
Table~\ref{tab:board-est} summarizes these calculations.
When compared to \aenc on the Arm board, we see a reduction of $27.95\%$ for Graviton.
However, when compared to \aprot, \aenc is still $\approx 2$ orders of magnitude slower. 
Note that, the performance of \aprot depends on the number of RSI calls which is a function of how many memory transfers occur. 
Whereas the performance of \aenc is impacted by the size of each memory transfer. 

\paragraph{Note.} Due to limited space, we defer implementation details and additional evaluation results to the extended version~\cite{acai-arxiv}.
\section{Related Work}
\paragraph{CPU TEEs.} 
Arm TrustZone provides secure world for sensitive computation~\cite{TZOS},
Intel SGX enclaves isolate user-level code~\cite{costan2016intel},
and AMD SEV protects confidential VMs~\cite{amd-sev}. 
Upcoming TEEs, such as Intel TDX~\cite{tdx} and Arm CCA~\cite{arm-da} offer  VM-level TEE abstractions.
Similarly, Sanctum~\cite{costan2016sanctum}, Keystone~\cite{keystone}, and Penglai~\cite{penglai} support TEEs on RISC-V platforms. 
These CPU-based TEEs do not support shared memory primitives between enclaves. 
Recent works offer memory sharing for RISC-V enclaves with PMP-based isolation~\cite{elasticlave, penglai}.
Further, several prior works create CPU-level TEEs using existing hardware mechanisms such as nested page tables and TrustZone address space controller~\cite{Overshadow,InkTag,brasser2019sanctuary}.
\codename re-purposes the hardware-enforced access-control and memory-sharing primitives
to create a protected region accessible to the realm VM and the device. 

\paragraph{PCIe Device TEEs.}
Graviton makes hardware modifications to support TEEs on GPUs~\cite{volos2018graviton} and Telekine makes them side-channel resistant~\cite{hunt2020telekine}. 
The same goal can be achieved on other devices~\cite{kang2021iceclave, zhao2021shef, dhar2022empowering} and  unmodified GPUs~\cite{jang2019HIX}. 
Nvidia Hopper H100 supports confidential computing, where the whole GPU can operate as a single isolated unit or can be viewed as multiple mutually untrusted isolated units called Multi-Instance GPUs (MIGs)~\cite{nvidia-azure}.
The device TEE hardware can perform line-rate encryption~\cite{nvidia-azure}. 
\codename requires such TEE-enabled devices to enforce it's invariants. 

\paragraph{Protecting Integrated Devices with CPU TEEs.}
Recent works extend TEEs from CPUs to integrated devices connected directly to the system bus.
Strongbox extends TrustZone-based isolation to integrated GPUs on Arm platforms while removing the driver from the TCB~\cite{strongbox}. 
Composite enclaves leverages PMP on RISC-V to create shared memory between enclaves and devices~\cite{schneider2022composite}.
Cure builds on RISC-V platforms and augments devices to expand the scope of CPU isolation to devices via bus-level enforcement~\cite{bahmani2020cure}. 
Similarly, CCA compliant integrated devices can connect to realm VMs if the device-side GPCs enforce access control. 
\codename's invariants are useful for these integrated devices, especially the SMMU protection and device binding. Addressing challenges in integrated mode may be easier---there is no dedicated device memory and the CPU protects the host DRAM.
Future work can investigate the security and compatibility of such devices using \codename's invariants.

\paragraph{TCB Reduction \& Verification.}
\codename does not aim to reduce TCB; it trusts the monitor, RMM, device driver, runtime, and guest kernel. 
Prior works remove the driver from the TCB for secure device communication with Arm TrustZone~\cite{strongbox, driverlets}. 
Shelter removes the RMM from the TCB by creating enclaves in the normal world instead of the realm world in Arm CCA~\cite{shelter}. 
Samsung  Islet, a Rust-based RMM, reduces potential memory safety issues~\cite{samsung-islet}. 
Further, the TrustZone firmware has been subject to formal verification~\cite{komodo, serval}. For Arm CCA, there are ongoing efforts to verify the RMM and the CCA firmware~\cite{via, tfx}.
Future works can use similar techniques to reduce \codename's TCB and formally verify it. 

\paragraph{\codename Takeways for Upcoming Platforms.}
Upcoming IOPMP in RISC-V~\cite{IOPMP}, PCIe-6 specification~\cite{pcie6}, its adoption as Intel TDX-Connect and AMD SEV-TIO are promising directions to expand \codename to other devices and CPU-TEEs~\cite{tdx-connect,amd-sev-tio,nvidia-azure}.
With these upcoming features, we hope that \codename serves as a guiding principle for future cloud deployments.
\codename invariants amend CCA's security specification to ensure that the CPU-side isolation extends to accelerators. 
Future works can use the same principle, but adapt it to specifics of the underlying isolation mechanisms: PMP and IOPMP in RISC-V, secure EPTs and IOMMU on Intel, or secure nested paging and IOMMU on AMD.
Adapting \codename to these systems will require revisiting challenges such as scalability and attestation, as well as compatibility with other hypervisor, kernel, and driver implementations.
\spacesave
\section{Conclusion}
\codename is the first system that demonstrates CCA-based confidential VMs 
to directly use accelerators as a first-class abstraction. 
\codename enforces strong isolation guarantees by extending existing CCA mechanisms to achieve security.

\section*{Acknowledgement}
We thank our shepherd and the anonymous reviewers for their feedback; 
Ben Fiedler, Ian Boschung, and Karin Holzhauser for fruitful discussions on Arm CCA; 
Valentin Ogier for adding encryption kernel support for Rodinia benchmarks and testing the QEMU support for Arm CCA. 

\balance 
\bibliographystyle{plain}
\bibliography{references} 
\appendix

\end{document}